\documentclass[longauth]{aa}

\usepackage{newtxtext,newtxmath}
\usepackage[T1]{fontenc}
\usepackage{color}
\usepackage{natbib,twoopt}
\usepackage[hyphenbreaks]{breakurl}
\usepackage[breaklinks]{hyperref}
\usepackage{graphicx}
\usepackage{upgreek}
\usepackage{xspace}
\usepackage[table]{xcolor}

\bibpunct{(}{)}{;}{a}{}{,}
\definecolor{cobalt}{rgb}{0.06, 0.2, 0.65}
\hypersetup{
  colorlinks,
  citecolor=cobalt,
  linkcolor=[rgb]{0.8, 0.2, 1.0},
  urlcolor=cobalt,
}

\newcommand{\lya}{Lyman-$\alpha$\xspace}
\newcommand{\lyaf}{Lyman-$\alpha$ forest\xspace}
\newcommand{\pcross}{$P_{\times}$\xspace}
\newcommand{\poned}{\ensuremath{P_{\rm 1D}}\xspace}
\newcommand{\xithreed}{\ensuremath{\xi_{\rm 3D}}\xspace}
\newcommand{\pthreed}{\ensuremath{P_{\rm 3D}}\xspace}

\newcommand{\forestflow}{\textsc{forestflow}\xspace}

\newcommand{\lacehc}{\textsc{training}\xspace}
\newcommand{\simseed}{\textsc{seed}\xspace}
\newcommand{\simigm}{\textsc{reionisation}\xspace}
\newcommand{\simcurved}{\textsc{curved}\xspace}
\newcommand{\simh}{\textsc{growth}\xspace}
\newcommand{\simnu}{\textsc{neutrinos}\xspace}

\newcommand{\simcentral}{\textsc{central}\xspace}

\newcommand{\mflux}{\ensuremath{\bar{F}}\xspace}
\newcommand{\iMpc}{\ensuremath{\,\mathrm{Mpc}^{-1}}}
\newcommand{\hMpc}{h^{-1}\,\mathrm{Mpc}}

\begin{document}

\title{ForestFlow: predicting the Lyman-$\alpha$ forest clustering from linear to nonlinear scales}
\titlerunning{ForestFlow: predicting Lyman-$\alpha$ forest clustering}

\author{
J.~Chaves-Montero\inst{\ref{inst0}}
\and
L.~Cabayol-Garcia\inst{\ref{inst0},\ref{inst1}}
\and
M.~Lokken\inst{\ref{inst0}}
\and
A.~Font-Ribera\inst{\ref{inst0}}
\and
J.~Aguilar\inst{\ref{inst2}}
\and
S.~Ahlen\inst{\ref{inst3}}
\and
D.~Bianchi\inst{\ref{inst19}}
\and
D.~Brooks\inst{\ref{inst23}}
\and
T.~Claybaugh\inst{\ref{inst2}}
\and
S.~Cole\inst{\ref{inst6}}
\and
A.~de la Macorra\inst{\ref{inst37}}
\and
S.~Ferraro\inst{\ref{inst2},\ref{inst43}}
\and
J.~E.~Forero-Romero\inst{\ref{inst52},\ref{inst53}}
\and
E.~Gaztañaga\inst{\ref{inst55},\ref{inst27}}
\and
S.~Gontcho A Gontcho\inst{\ref{inst2}}
\and
G.~Gutierrez\inst{\ref{inst25}}
\and
K.~Honscheid\inst{\ref{inst33},\ref{inst46},\ref{inst47}}
\and
R.~Kehoe\inst{\ref{inst66}}
\and
D.~Kirkby\inst{\ref{inst14}}
\and
A.~Kremin\inst{\ref{inst2}}
\and
A.~Lambert\inst{\ref{inst2}}
\and
M.~Landriau\inst{\ref{inst2}}
\and
M.~Manera\inst{\ref{inst75},\ref{inst0}}
\and
P.~Martini\inst{\ref{inst33},\ref{inst65},\ref{inst47}}
\and
R.~Miquel\inst{\ref{inst77},\ref{inst0}}
\and
A.~Muñoz-Gutiérrez\inst{\ref{inst37}}
\and
G.~Niz\inst{\ref{inst36},\ref{inst11}}
\and
I.~P\'erez-R\`afols\inst{\ref{inst84}}
\and
G.~Rossi\inst{\ref{inst88}}
\and
E.~Sanchez\inst{\ref{inst38}}
\and
M.~Schubnell\inst{\ref{inst7}}
\and
D.~Sprayberry\inst{\ref{inst21}}
\and
G.~Tarl\'{e}\inst{\ref{inst7}}
\and
B.~A.~Weaver\inst{\ref{inst21}}
}

\institute{
Institut de F\'{\i}sica d'Altes Energies (IFAE), The Barcelona Institute of Science and Technology, 08193 Bellaterra (Barcelona), Spain \label{inst0}\\
\email{jchaves@ifae.es}
\and
Port d'Informaci\'{o} Cient\'{i}fica, Campus UAB, C. Albareda s/n, 08193 Bellaterra (Barcelona), Spain \label{inst1}
\and
Lawrence Berkeley National Laboratory, 1 Cyclotron Road, Berkeley, CA 94720, USA \label{inst2}
\and
Physics Dept., Boston University, 590 Commonwealth Avenue, Boston, MA 02215, USA \label{inst3}
\and
Dipartimento di Fisica ``Aldo Pontremoli'', Universit\`a degli Studi di Milano, Via Celoria 16, I-20133 Milano, Italy \label{inst19}
\and
Department of Physics \& Astronomy, University College London, Gower Street, London, WC1E 6BT, UK \label{inst23}
\and
Institute for Computational Cosmology, Department of Physics, Durham University, South Road, Durham DH1 3LE, UK \label{inst6}
\and
Instituto de F\'{\i}sica, Universidad Nacional Aut\'{o}noma de M\'{e}xico,  Cd. de M\'{e}xico  C.P. 04510,  M\'{e}xico \label{inst37}
\and
University of California, Berkeley, 110 Sproul Hall \#5800 Berkeley, CA 94720, USA \label{inst43}
\and
Departamento de F\'isica, Universidad de los Andes, Cra. 1 No. 18A-10, Edificio Ip, CP 111711, Bogot\'a, Colombia \label{inst52}
\and
Observatorio Astron\'omico, Universidad de los Andes, Cra. 1 No. 18A-10, Edificio H, CP 111711 Bogot\'a, Colombia \label{inst53}
\and
Institut d'Estudis Espacials de Catalunya (IEEC), 08034 Barcelona, Spain \label{inst55}
\and
Institute of Cosmology and Gravitation, University of Portsmouth, Dennis Sciama Building, Portsmouth, PO1 3FX, UK \label{inst27}
\and
Institute of Space Sciences, ICE-CSIC, Campus UAB, Carrer de Can Magrans s/n, 08913 Bellaterra, Barcelona, Spain \label{inst56}
\and
Fermi National Accelerator Laboratory, PO Box 500, Batavia, IL 60510, USA \label{inst25}
\and
Center for Cosmology and AstroParticle Physics, The Ohio State University, 191 West Woodruff Avenue, Columbus, OH 43210, USA \label{inst33}
\and
Department of Physics, The Ohio State University, 191 West Woodruff Avenue, Columbus, OH 43210, USA \label{inst46}
\and
The Ohio State University, Columbus, 43210 OH, USA \label{inst47}
\and
Department of Physics, Southern Methodist University, 3215 Daniel Avenue, Dallas, TX 75275, USA \label{inst66}
\and
Department of Physics and Astronomy, University of California, Irvine, 92697, USA \label{inst14}
\and
Departament de F\'{i}sica, Serra H\'{u}nter, Universitat Aut\`{o}noma de Barcelona, 08193 Bellaterra (Barcelona), Spain \label{inst75}
\and
Department of Astronomy, The Ohio State University, 4055 McPherson Laboratory, 140 W 18th Avenue, Columbus, OH 43210, USA \label{inst65}
\and
Instituci\'{o} Catalana de Recerca i Estudis Avan\c{c}ats, Passeig de Llu\'{\i}s Companys, 23, 08010 Barcelona, Spain \label{inst77}
\and
Departamento de F\'{i}sica, Universidad de Guanajuato - DCI, C.P. 37150, Leon, Guanajuato, M\'{e}xico \label{inst36}
\and
Instituto Avanzado de Cosmolog\'{\i}a A.~C., San Marcos 11 - Atenas 202. Magdalena Contreras, 10720. Ciudad de M\'{e}xico, M\'{e}xico \label{inst11}
\and
Departament de F\'isica, EEBE, Universitat Polit\`ecnica de Catalunya, c/Eduard Maristany 10, 08930 Barcelona, Spain \label{inst84}
\and
Department of Physics and Astronomy, Sejong University, Seoul, 143-747, Korea \label{inst88}
\and
CIEMAT, Avenida Complutense 40, E-28040 Madrid, Spain \label{inst38}
\and
Department of Physics, University of Michigan, Ann Arbor, MI 48109, USA \label{inst7}
\and
NSF NOIRLab, 950 N. Cherry Ave., Tucson, AZ 85719, USA \label{inst21}
}

\date{Received 2024; accepted XXX}


\abstract{
On large scales, the Lyman-$\alpha$ forest provides insights into the expansion history of the Universe, while on small scales, it imposes strict constraints on the growth history, the nature of dark matter, and the sum of neutrino masses. This work introduces ForestFlow, a novel framework that bridges the gap between large- and small-scale analyses, which have traditionally relied on distinct modeling approaches. Using conditional normalizing flows, ForestFlow predicts the two Lyman-$\alpha$ linear biases ($b_\delta$ and $b_\eta$) and six parameters describing small-scale deviations of the three-dimensional flux power spectrum ($P_\mathrm{3D}$) from linear theory as a function of cosmology and intergalactic medium physics. These are then combined with a Boltzmann solver to make consistent predictions, from arbitrarily large scales down to the nonlinear regime, for $P_\mathrm{3D}$ and any other statistics derived from it. Trained on a suite of 30 fixed-and-paired cosmological hydrodynamical simulations spanning redshifts from $z=2$ to 4.5, ForestFlow achieves 3 and 1.5\% precision in describing $P_\mathrm{3D}$ and the one-dimensional flux power spectrum ($P_\mathrm{1D}$) from linear scales to $k=5\,\mathrm{Mpc}^{-1}$ and $k_\parallel=4\,\mathrm{Mpc}^{-1}$, respectively. Thanks to its conditional parameterization, ForestFlow shows similar performance for ionization histories and two $\Lambda$CDM model extensions --- massive neutrinos and curvature --- even though none of these are included in the training set. This framework will enable full-scale cosmological analyses of Lyman-$\alpha$ forest measurements from the DESI survey.
}

\maketitle



\section{Introduction}

The \lyaf refers to absorption lines in the spectra of high-redshift quasars resulting from \lya absorption by neutral hydrogen in the intergalactic medium \citep[IGM; for a review, see][]{mcquinn2016EvolutionIntergalacticMedium}. Even though quasars can be observed at very high redshifts with relatively short exposure times, the scarcity of these sources limits their direct use for precision cosmology. Conversely, \lyaf measurements from a single quasar spectrum provide information about density fluctuations over hundreds of megaparsecs along the line of sight, making this observable an excellent tracer of large-scale structure at high redshifts.

Cosmological analyses of the \lyaf rely on either three-dimensional correlations of the \lya transmission field \citep[\xithreed; e.g.,][]{slosar2011LymanaForestThree} or correlations along the line-of-sight of each quasar \citep[i.e., the one-dimensional flux power spectrum \poned,][]{emuparam_croft, mcdonald2000ObservedProbabilityDistribution}. The first analyses set constraints on the expansion history of the Universe by measuring baryonic acoustic oscillations \citep[BAO; e.g.,][]{busca2013LyaBAODR9, slosar2013MeasurementBaryonAcoustica, dumasdesbourboux2020CompletedSDSSIVExtended}, for which linear or perturbation theory is accurate enough. On the other hand, \poned analyses measure the small-scale amplitude and slope of the linear power spectrum \citep[e.g.,][]{emuparam_croft, mcdonald2000ObservedProbabilityDistribution,zaldarriaga2001ConstraintsLyaForest, viel2004ConstraintsPrimordialPower, mcdonald2005LinearTheoryPower}, the nature of dark matter \citep[e.g.,][]{2006PhRvL..97s1303S, 2013PhRvD..88d3502V, irsic2017FirstConstraintsFuzzy, palanque-delabrouille2020HintsNeutrinoBounds, gpemucosmo_rogers, irsic2024}, the thermal history of the IGM \citep[e.g.,][]{viel2006CosmologicalAstrophysicalParameters, bolton2008PossibleEvidenceInverted, lee2015IGMConstraintsSDSSIII, emugp_Walther2019, Boera2019, Gaikwad2020, Gaikwad2021} and the reionization history of the Universe \citep[see the reviews][]{, meiksin2009PhysicsIntergalacticMedium, mcquinn2016EvolutionIntergalacticMedium}. In combination with cosmic microwave background constraints, \poned analyses also set tight constraints on the sum of neutrino masses and the running of the spectral index \citep[e.g.,][]{spergel2003FirstYearWilkinsonMicrowave, verde2003FirstYearWilkinsonMicrowave, viel2004ConstraintsPrimordialPower, seljak2005CosmologicalParameterAnalysis, seljak2006CosmologicalParametersCombining, palanque-delabrouille2015ConstraintNeutrinoMasses, palanque-delabrouille2020HintsNeutrinoBounds}.

Unlike \xithreed studies, \poned analyses go deep into the nonlinear regime and require time-demanding hydrodynamical simulations \citep[e.g.,][]{hydro_Cen1994, hydro_Miralda1996, hydro_Meiksin2001, hydro_Lukic2015, bolton2017SherwoodSimulationSuite, hydro_Walther2021, hydro_Chabanier2023, Puchwein2023, bird2023PRIYANewSuite}. Naive analyses would demand running millions of hydrodynamical simulations, which is currently unfeasible. Rather, the preferred solution is constructing fast surrogate models that make precise predictions across the input parameter space using simulation measurements as the training set. The main advantage of these surrogate models, known as emulators, is reducing the number of simulations required for Bayesian inference from millions to dozens or hundreds. In the context of \lyaf studies, the first \poned emulators involved simple linear interpolation \citep{mcdonald2006LyaForestPowera} and progressively moved toward using Gaussian processes \citep[GPs;][]{sacks1989DesignAnalysisComputer, mackay1998introduction} and neural networks \citep[NNs;][]{mcculloch1943logical}; for instance, \citet{emugp_bird2019}, \citet{emugp_rogers2019}, \citet{emugp_Walther2019}, \citet{Pedersen2021}, \citet{emugp_Takhtaganov2021}, \citet{emugp_rogers2021}, \citet{gpemu:P1DFernandez2022}, \citet{bird2023PRIYANewSuite}, \citet{ennemu:P1DMolaro2023}, and \citet{cabayol-garcia2023NeuralNetworkEmulator}.

The primary purpose of this work is to provide consistent predictions for \lyaf clustering statistics from linear to nonlinear scales. There are three main approaches to achieve this. The first relies on perturbation theory \citep[e.g.,][]{GivansHirata2020, ChenVlah2021, Ivanov2024}, which delivers precise predictions on perturbative scales at the cost of marginalizing over a large number of free parameters. The second involves emulating power spectrum modes measured from a suite of cosmological hydrodynamical simulations, which provides precise predictions from quasilinear to nonlinear scales. The main limitation of this approach is that accessing the largest scales used in BAO analyses, $r\simeq300$ Mpc, would require hydrodynamical simulations at least three times larger than this scale \citep[e.g.,][]{Angulo2008}, which is currently unfeasible due to the computational demands of these simulations. Nonetheless, there are recent strides in this direction using approximated methods \citep[e.g.,][]{jacobus2023ReconstructingLyaFields}.

Instead, we adopt a third approach: emulating the best-fitting parameters of a physically motivated \lya clustering model to measurements from a suite of cosmological hydrodynamical simulations as a function of cosmology and IGM physics. In what follows, we refer to this strategy as \forestflow\footnote{Publicly available at \url{https://github.com/igmhub/ForestFlow}.}. In this work, we use conditional normalizing flows \citep[cNF;][]{Winkler2019, cNF_Papamakarios} to emulate the two \lya linear biases ($b_\delta$ and $b_\eta$), which completely determine the large-scale behavior of \pthreed in conjunction with the linear power spectrum, along with six parameters capturing small-scale deviations of \pthreed from linear theory as a function of six parameters capturing the cosmological and IGM dependence of \lya clustering \citep{Pedersen2021}. We show that this strategy enables precise \pthreed predictions from nonlinear scales to arbitrarily large (linear) scales even when training on a suite of simulations with moderate size \citep{Pedersen2021, cabayol-garcia2023NeuralNetworkEmulator}. \forestflow also allows for the prediction of any statistic derived from \pthreed without requiring interpolation or extrapolation. For instance, we can compute \xithreed by taking the Fourier transform of \pthreed or derive \poned by integrating its perpendicular modes
\begin{equation}
    \label{eq:p1d}
    \poned(k_\parallel)=(2\uppi)^{-1}\int_0^\infty \mathrm{d} k_\perp\, k_\perp\, \pthreed(k_\parallel,\, k_\perp),
\end{equation}
where $k_\parallel$ and $k_\perp$ indicate parallel and perpendicular modes, respectively.

The release of \forestflow is quite timely for BAO and \poned analyses of the ongoing Dark Energy Spectroscopic Instrument survey \citep[DESI;][]{DESI_collab2016}, which will quadruple the number of line-of-sights observed by first the Baryon Oscillation Spectroscopic Survey \citep[BOSS;][]{boss_dawson2013} and its extension \citep[eBOSS;][]{eboss_dawson2016}. DESI has already proven the constraining power of \lya studies by measuring the isotropic BAO scale with $\simeq1\%$ precision from the Data Release 1 \citep{desicollaboration2024DESI2024IV} and \poned at nine redshift bins with a precision of a few percent from the Early Data Release \citep{ravoux2023DarkEnergySpectroscopica, karacayli2024Optimal1DLy}. In addition to being used for BAO and \poned studies, \forestflow has the potential to extend toward nonlinear scales the current full-shape analyses of \xithreed \citep{cuceu2023ConstraintsCosmicExpansion, 2023MNRAS.518.2567G} and \pthreed \citep{fontribera2018HowEstimate3D, Belsunce2024eBOSS, Horowitz2024}, and can be used to interpret alternative three-dimensional statistics \citep{hui1999GeometricalTestCosmological, fontribera2018HowEstimate3D, Karim2023}.

The structure of this paper is as follows: we describe the strategy adopted by \forestflow, the input data for the cNF, and its architecture in Sect.~\ref{sec:strategy}, \ref{sec:input}, and \ref{sec:forestflow}, respectively. Next, we assess the performance of \forestflow in Sect.~\ref{sec:results} and highlight some novel analyses enabled by this framework in Sect.~\ref{sec:discussion}. Finally, we summarize the main findings and conclude in Sect.~\ref{sec:conclusions}. Throughout this paper, all statistics and distances are in comoving units.


\section{Strategy adopted by \forestflow}
\label{sec:strategy}

In \forestflow, we emulate the parameters of a physically motivated parametric model for \lyaf clustering as a function of parameters that capture the influence of cosmology and IGM physics on this observable. This section begins with an overview of the physically motivated model, followed by an introduction to the parameters used to characterize the dependence of \lyaf clustering on cosmology and IGM physics.


\subsection{Parametric model for \pthreed}
\label{sec:strategy_model}

\begin{figure}
\includegraphics[width=\columnwidth]{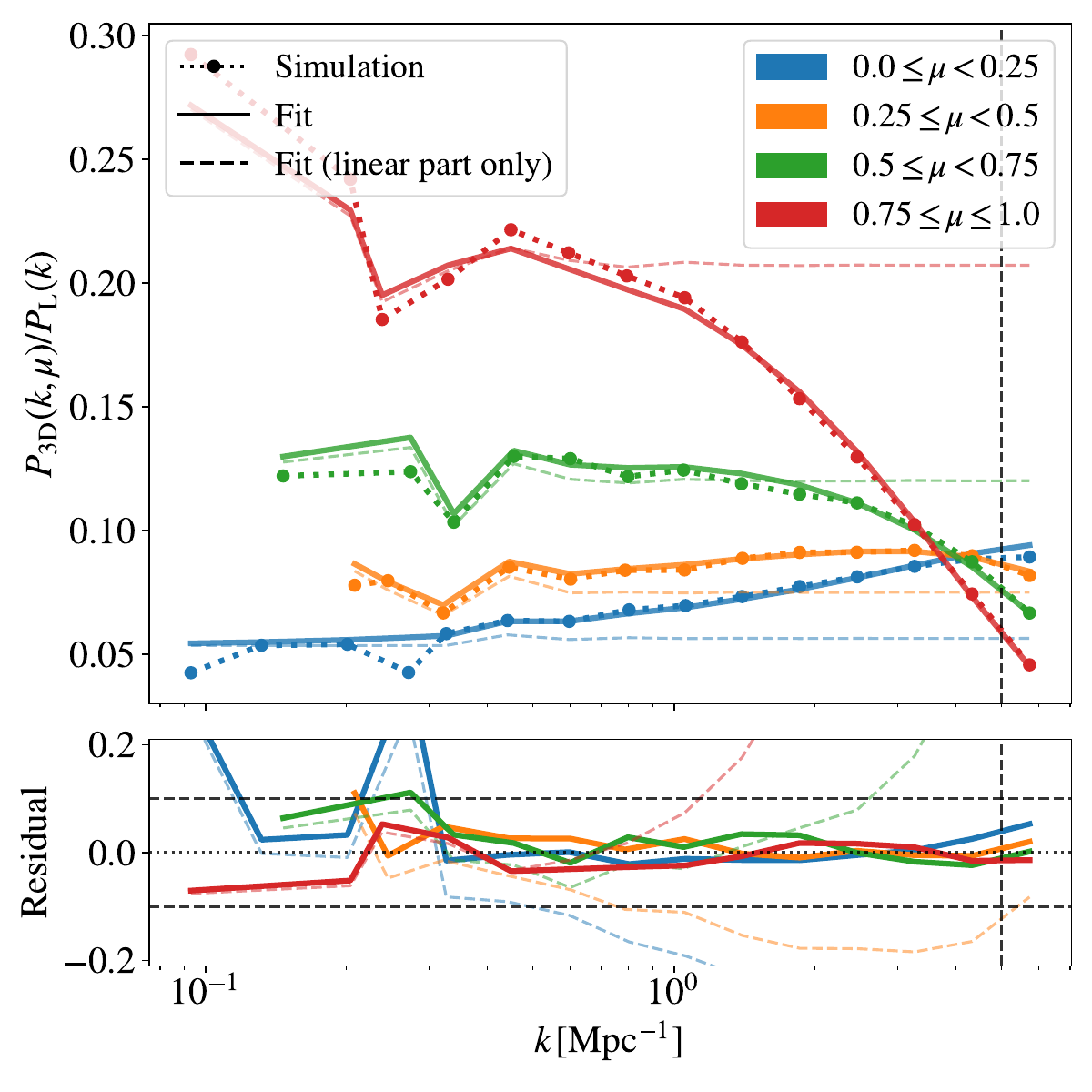}
\centering
\caption{Accuracy of the \pthreed model (see Eqs.~\ref{eq:p3d_model} and \ref{eq:dnl}) in reproducing measurements from the \simcentral simulation at $z=3$. In the top panel, dotted and solid lines show the ratio of simulation measurements and model predictions relative to the linear power spectrum, respectively. Dashed lines do so for the linear part of the best-fitting model ($D_\mathrm{NL}=1$). Line colors correspond to different $\mu$ wedges, and vertical dashed lines mark the minimum scale used for computing the best-fitting model, $k=5\iMpc$. The bottom panel displays the relative difference between the best-fitting model and simulation measurements. The overall accuracy of the model is 2\% on scales in which simulation measurements are not strongly affected by cosmic variance ($k>0.5\iMpc$; see text).
}
\label{fig:arinyo}
\end{figure}

We can express fluctuations in the \lyaf flux as $\delta_F(\mathbf{s}) = \bar{F}^{-1}(\mathbf{s}) F(\mathbf{s})-1$, where $F=\exp(-\tau)$ and $\bar{F}$ are the transmitted flux fraction and its mean, respectively, $\tau$ is the optical depth to \lya absorption, and $\mathbf{s}$ is the redshift-space coordinate. On linear scales, these fluctuations depend upon the matter field as follows \citep[e.g.,][]{mcdonald2003MeasurementCosmologicalGeometry} 
\begin{equation}
    \delta_F=b_\delta\, \delta + b_\eta\, \eta,
\end{equation}
where $\delta$ refers to matter density fluctuations, $\eta=-(a\,H)^{-1}(\partial v_\mathrm{r}/\partial r)$ stands for the dimensionless line-of-sight gradient of radial peculiar velocities, $a$ is the cosmological expansion factor, $H$ is the Hubble expansion factor, $v_\mathrm{r}$ is the radial velocity, and $r$ stands for the radial comoving coordinate. The linear bias coefficients $b_\delta$ and $b_\eta$ capture the response of $\delta_F$ to large-scale fluctuations in the $\delta$ and $\eta$ fields, respectively. 

Following \citet{mcdonald2003MeasurementCosmologicalGeometry}, we decompose the three-dimensional power spectrum of $\delta_F$ into three terms
\begin{equation}
    \label{eq:p3d_model}
    \pthreed(k,\,\mu) = (b_\delta + b_\eta\, f\, \mu^2)^2 D_\mathrm{NL}(k,\,\mu) P_\mathrm{lin}(k),
\end{equation}
where $f=\mathrm{d}\log G/\mathrm{d}\log a$ is the logarithmic derivative of the growth factor $G$, $\mu=k^{-1}k_\parallel$ is the cosine of the angle between the Fourier mode and the line of sight, $(b_\delta + b_\eta\, f\, \mu^2)^2$ accounts for both linear biasing and large-scale redshift space distortions \citep{kaiser1987ClusteringRealSpace, mcdonald2000ObservedProbabilityDistribution}, $P_\mathrm{lin}$ is the linear matter power spectrum\footnote{This is the linear power spectrum of cold dark matter and baryons even for cosmologies with massive neutrinos.}, and $D_\mathrm{NL}$ is a physically motivated parametric correction accounting for the nonlinear growth of the density field, nonlinear peculiar velocities, thermal broadening, and pressure.

The large-scale behavior of \pthreed is set by the bias coefficients $b_\delta$ and $b_\eta$ together with the linear power spectrum, and the latter can be computed using a Boltzmann solver \citep[e.g.,][]{lewis2000EfficientComputationCosmic, lesgourgues2011CosmicLinearAnisotropy}. Therefore, the emulation of the two \lya linear biases enables predicting \pthreed on arbitrarily large (linear) scales\footnote{Aside from nonlinear effects affecting the position and damping of BAO.}. In contrast with direct emulation of power spectrum modes, this approach only requires simulations large enough for measuring the two \lya linear biases precisely.

Predicting \pthreed on small scales is more challenging than on large scales due to the variety of effects influencing this statistic in the nonlinear regime \citep[e.g.,][]{mcdonald2003MeasurementCosmologicalGeometry}. In this work, we capture these small-scale effects using the physically motivated \citet{arinyo-i-prats2015NonlinearPowerSpectrum} parameterization
\begin{equation}
    \label{eq:dnl}
    D_\mathrm{NL} = \exp \left\{\left(q_1 \Delta^2 + q_2 \Delta^4\right) \left[1-\left(\frac{k}{k_\mathrm{v}}\right)^{a_\mathrm{v}} \mu^{b_\mathrm{v}}\right] - \left(\frac{k}{k_\mathrm{p}}\right)^2 \right\},
\end{equation}
where $\Delta^2(k)\equiv(2\uppi^2)^{-1} k^3 P_\mathrm{lin}(k)$ is the dimensionless linear matter power spectrum and the free parameters $k_\mathrm{v}$ and $k_\mathrm{p}$ are in $\iMpc$ units throughout this work. The terms involving $\{q_1,\,q_2\}$, $\{k_\mathrm{v},\, a_\mathrm{v},\, b_\mathrm{v}\}$, and $\{k_\mathrm{p}\}$ account for nonlinear growth, peculiar velocities and thermal broadening, and gas pressure, respectively. The expression above does not account for a shot noise term \citep[e.g.,][]{irsicmcquinn2018}. While \citet{givans2022NonlinearitiesLymanAlpha} successfully described \pthreed and \poned measurements down to highly nonlinear scales using this formulation, it is possible that the shot noise contribution was implicitly absorbed into their fit through free parameters representing other effects. A more detailed investigation of shot noise is deferred to future work.

In the top panel of Fig.~\ref{fig:arinyo}, dotted lines show the ratio of measurements from the \simcentral simulation at $z=3$ and the linear power spectrum, while solid lines do so for the best-fitting model to these measurements (Eqs.~\ref{eq:p3d_model} and \ref{eq:dnl}) and the linear power spectrum. See Sect.~\ref{sec:input} for details about this simulation and the fitting procedure. The dashed lines depict the results for the best-fitting model when setting $D_\mathrm{NL}=1$ after carrying out the fit; in other words, the behavior of the best-fitting model on linear scales. We can readily see that nonlinear growth isotropically increases the power with growing $k$, while peculiar velocities and thermal broadening suppress the power of parallel modes as $k$ increases. On even smaller scales, pressure takes over and causes an isotropic suppression. Nonlinear growth modifies the perpendicular power relative to linear theory by 10\% for scales as large as $k=0.5\iMpc$, indicating that small-scale corrections are important for most of the scales sampled by our simulations. Nevertheless, in Appendix~\ref{sec:cosmic_variance}, we show that we can measure the two \lya linear biases with percent precision from these simulations. Deviations from linear theory are less pronounced down to smaller scales for modes with $\mu\simeq0.5$ because nonlinear growth and the combination of peculiar velocities and thermal broadening tend to cancel each other out. As we can see, the parametric model achieves an average accuracy of 2\% for $k>0.5\iMpc$, supporting the validity of Eq.\ref{eq:dnl} for capturing small-scale deviations from linear theory.

On the largest scales, we find strong variations between consecutive $k$-bins for the same $\mu$-wedge. Some of these oscillations are driven by differences in the average value of $\mu$ between consecutive bins due to the limited number of modes entering each bin on large scales. To ensure an accurate comparison between simulation measurements and model predictions, we individually evaluate the \pthreed model for all the modes within each $k-\mu$ bin from our simulation boxes. We then calculate the mean of the resulting distribution and assign this mean value to the bin, thereby mirroring the approach used to compute \pthreed measurements from the simulations. This process is also important on small scales, where the number of modes increases rapidly with $k$. Throughout this work, we adopt this approach to make predictions from the \pthreed model.

Using this approach to generate theoretical predictions, the best-fitting model successfully captures most of the aforementioned large-scale oscillations. However, a fluctuation at $k\simeq0.25\iMpc$ in the $0<\mu<0.25$ wedge remains unaccounted for by the model. The negligible differences between model predictions and simulation measurements in the adjacent bins suggest that this fluctuation is likely due to cosmic variance. We assess the impact of this source of uncertainty on \pthreed in Appendix~\ref{sec:cosmic_variance}, finding that it can induce fluctuations of up to 10\% on scales $k<0.5\iMpc$. As a result, cosmic variance limits our ability to evaluate the model's performance on the largest scales shown. However, this does not necessarily reflect reduced model accuracy but rather highlights the limitations of using our simulations for validating the model. Proper validation on large scales would require either a larger simulation or multiple simulations with different initial conditions.


\subsection{Input and output parameters}
\label{sec:strategy_params}

In addition to the density and velocity fields, the \lyaf depends upon the ionization and thermal state of the IGM \citep[e.g.,][]{mcdonald2003MeasurementCosmologicalGeometry}. Following \citet{Pedersen2021}, we use six parameters to describe the dependency of this observable with cosmology and IGM physics:

\begin{itemize}
    \item Amplitude and slope of the linear matter power spectrum on small scales. We define the amplitude ($\Delta^2_\mathrm{p}$) and slope ($n_\mathrm{p}$) as
    \begin{align}
        \label{eq:amplitude}
        & \Delta_\mathrm{p}^2(z) = (2\pi^2)^{-1} k^3 P_{\rm lin} (k_\mathrm{p}, z),\\
        \label{eq:slope}
        & n_\mathrm{p}(z) = \left(\mathrm{d}\log P_{\rm lin} / \mathrm{d}\log k\right)\mid_{k = k_\mathrm{p}},
    \end{align}
    where we use $k_\mathrm{p} = 0.7\,\mathrm{Mpc}^{-1}$ as the pivot scale because it is at the center of the range of interest for DESI small-scale studies. These parameters capture multiple physical effects modifying the linear power spectrum on small scales \citep[see][for a detailed discussion]{Pedersen2021}, including cosmological parameters such as the amplitude ($A_\mathrm{s}$) and slope ($n_\mathrm{s}$) of the primordial power spectrum, the Hubble parameter, and the matter density ($\Omega_\mathrm{M}$), or $\Lambda$CDM extensions such as curvature and massive neutrinos. The advantage of using this parameterization rather than $\Lambda$CDM parameters is twofold. First, it reduces the dimensionality of the input to the cNF (see Sect.~\ref{sec:forestflow}), which decreases the number of simulations required for precise training. Second, the resulting emulator has the potential for making precise predictions for variations in cosmological parameters and $\Lambda$CDM extensions not considered in the training set \citep[][]{Pedersen2021, pedersen2023CompressingCosmologicalInformation, cabayol-garcia2023NeuralNetworkEmulator}. Note that we do not consider cosmological parameters capturing changes in the growth rate or expansion history because the \lyaf probes cosmic times during which the universe is practically Einstein de-Sitter, and both vary very little with cosmology in this regime.

    \item Mean transmitted flux fraction. The mean transmitted flux fraction (\mflux) depends on the intensity of the cosmic ionizing background and evolves strongly with redshift. One of the advantages of using this parameter is that it encodes the majority of the redshift dependence of the signal, serving as a proxy for cosmic time.

    \item Amplitude and slope of the temperature-density relation. The thermal state of the IGM can be approximated by a power law on the densities probed by the \lyaf \citep{hydro_Lukic2015}: $T_0\Delta_\mathrm{b}^{\gamma-1}$, where $\Delta_\mathrm{b}$ is the baryon overdensity, $T_0$ is the gas temperature at mean density, and $\gamma-1$ is the slope of the relation. These parameters influence the ionization of the IGM, which is captured by \mflux, and the thermal motion of gas particles, which causes Doppler broadening that suppresses the parallel power. Instead of using $T_0$ as an emulator parameter, we follow \citet{Pedersen2021} and use the thermal broadening scale in comoving units. First, we express the thermal broadening in velocity units as $\tilde{\sigma}_\mathrm{T} = 9.1 (T_0[\mathrm{K}]/10^4)^{1/2}$, and then we convert it to comoving units, $\sigma_\mathrm{T}=\tilde{\sigma}_\mathrm{T}(1+z) H^{-1}$.

    \item Pressure smoothing scale. Gas pressure supports baryons on small scales, leading to a strong isotropic power suppression in this regime that depends upon the entire thermal history of the gas \citep{gnedin1998ProbingUniverseLyalpha}. We parameterize this effect using the pressure smoothing scale in units of comoving Mpc$^{-1}$, $k_\mathrm{F}$ \citep[see][for more details]{Pedersen2021}.
\end{itemize}

In summary, our cNF predicts the eight free parameters of the physically motivated model for \lya clustering introduced by Eqs.~\ref{eq:p3d_model} and \ref{eq:dnl}, $\mathbf{y}=\{b_\delta,\, b_\eta,\, q_1,\, q_2,\, k_\mathrm{v},\, a_\mathrm{v},\, b_\mathrm{v}, \, k_\mathrm{p}\}$, as a function of the aforementioned six parameters capturing the cosmological and IGM dependence of the \lyaf, $\mathbf{x}=\{\Delta_\mathrm{p}^2,\, n_\mathrm{p},\, \mflux,\, \sigma_\mathrm{T},\, \gamma,\, k_\mathrm{F}\}$.


\section{Training and testing set}
\label{sec:input}

In this section, we describe how we generated the training and testing data for the cNF described in Sect.~\ref{sec:forestflow}. In Sect.~\ref{sec:input_sims}, we present a suite of cosmological hydrodynamical simulations from which we generated mock \lyaf measurements, and we detail our approach for extracting \pthreed and \poned measurements from these simulations in Sect.~\ref{sec:input_extract_lya}. In Sect.~\ref{sec:input_fitting}, we compute the best-fitting parameters of the model introduced by Eqs.~\ref{eq:p3d_model} and \ref{eq:dnl} to measurements of these statistics, and we evaluate the performance of the fits in Sect.~\ref{sec:input_precision}.


\subsection{Simulations}
\label{sec:input_sims}

We extracted \lyaf simulated measurements from a suite of simulations run with \textsc{mp-gadget}\footnote{\url{https://github.com/MP-Gadget/MP-Gadget/}} \citep{feng2018MpGadgetMpGadgetTag, emugp_bird2019}, a massively scalable version of the cosmological structure formation code \textsc{gadget-3} \citep[last described in][]{Gadget_Springel}. This suite was first presented and used in \citet{Pedersen2021}; we briefly describe it next. Each simulation tracked the evolution of $768^3$ dark matter and baryon particles from $z=99$ to $z=2$ inside a box of $L = 67.5$ Mpc on a side, producing as output 11 snapshots uniformly spaced in redshift between $z=4.5$ and 2. This configuration ensures convergence for \poned measurements down to $k_\parallel=4\iMpc$ (the smallest scale used in this work) at $z=2$ and less than 10\% errors for this scale at $z=4$ \citep[see][for more details]{bolton2017SherwoodSimulationSuite}. On the other hand, this configuration may cause non-negligible biases for \pthreed at high redshift \citep{hydro_Lukic2015}.

Two realizations were run for each combination of cosmological and astrophysical parameters using the ``fixed-and-paired'' technique \citep{angulo2016CosmologicalNbodySimulations, pontzen2016InvertedInitialConditions}, which significantly reduces cosmic variance for multiple observables, including the \lyaf \citep{fixedpaired_Villaescusa, anderson2019CosmologicalHydrodynamicSimulations}. The initial conditions were generated using the following configuration of \textsc{mp-genic} \citep{Bird2020}: initial displacements produced using the Zel'dovich approximation and baryons and dark matter initialized on an offset grid using species-specific transfer functions. Some studies have suggested that this configuration might lead to incorrect evolution of linear modes \citep{Bird2020}. However, in a recent study, \citet{Khan2024} showed that variations in the specific setting of \textsc{mp-genic} initial conditions have a minimal impact on \poned measurements across the range of redshifts and scales used in this work.

To increase computational efficiency, the simulations utilized a simplified prescription for star formation that turns regions of baryon overdensity $\Delta_\mathrm{b}>1000$ and temperature $T<10^5$ K into collisionless stars \citep[e.g.,][]{viel2004ConstraintsPrimordialPower}, implemented a spatially uniform ultraviolet background \citep{haardt2012RadiativeTransferClumpya}, and did not consider active galactic nuclei (AGN) feedback \citep[e.g.,][]{chabanier2020ImpactAGNFeedback}. These approximations are justified because we focus on emulating the \lyaf in the absence of astrophysical contaminants like AGN feedback, damped Lyman-alpha absorbers (DLAs), or metal absorbers, and we will model these before comparing our predictions with observational measurements \citep[e.g.,][]{mcdonald2005LinearTheoryPower, palanque-delabrouille2015ConstraintNeutrinoMasses, palanque-delabrouille2020HintsNeutrinoBounds}.

In Sect.~\ref{sec:forestflow}, we train a cNF using data from 30 fixed-and-paired simulations from the previous suite, covering combinations of cosmological and astrophysical parameters selected via a Latin hypercube sampling method \citep{mckay1979ComparisonThreeMethods}. Hereafter, we refer to these as \lacehc simulations. The Latin hypercube spans the parameters $\{\Delta^2_\mathrm{p}(z=3),\, n_\mathrm{p}(z=3),\, z_\mathrm{H},\, H_\mathrm{A},\, H_\mathrm{S}\}$, where we use $z=3$ because it is approximately at the center of the range of interest for DESI studies \citep{ravoux2023DarkEnergySpectroscopica, karacayli2024Optimal1DLy}, $z_\mathrm{H}$ is the midpoint of hydrogen reionization, and the last two parameters rescale the total photoheating rate $\epsilon_0$ as $\epsilon = H_\mathrm{A} \Delta_\mathrm{b}^{H_\mathrm{S}} \epsilon_0$ \citep{onorbe2017SelfconsistentModelingReionization}. Cosmological parameters were generated within the ranges $\Delta^2_\mathrm{p}(z=3) \in [0.25,\, 0.45]$, $n_\mathrm{p}(z=3) \in [-2.35,\, -2.25]$ by exploring values of the amplitude and slope of the primordial power spectrum within the intervals $A_\mathrm{s} \in [1.35,\, 2.71]\times 10^{-9}$ and $n_\mathrm{s} \in [0.92,\, 1.02]$. Any other $\Lambda$CDM parameter was held fixed to values approximately following \citet{planckcollaboration2020Planck2018Resultsa}: dimensionless Hubble parameter $h=0.67$, physical cold dark matter density $\Omega_\mathrm{c} h^2=0.12$, and physical baryon density $\Omega_\mathrm{b} h^2=0.022$. As for the IGM parameters, these explored the ranges $z_\mathrm{H}\in[5.5,\,15]$, $H_\mathrm{A}\in[0.5,\,1.5]$, and $H_\mathrm{S}\in[0.5,\,1.5]$. All simulation pairs share the same set of initial Fourier phases, making their \pthreed and \poned subject to the same large-scale noise pattern.

We evaluated different aspects of the emulation strategy using six fixed-and-paired simulations with cosmological and astrophysical parameters not considered in the \lacehc simulations:
\begin{itemize}
    \item The \simcentral simulation uses cosmological and astrophysical parameters at the center of the \lacehc parameter space: $A_\mathrm{s}=2.01\times10^{-9}$, $n_\mathrm{s}=0.97$, $z_\mathrm{H}=10.5$, $H_\mathrm{A}=1$, and $H_\mathrm{S}=1$. We use this simulation for an out-of-sample test to evaluate the performance of \forestflow under optimal conditions, recognizing that the accuracy of machine-learning models generally declines near the boundaries of the convex hull defined by the training set.
    
    \item The \simseed simulation uses the same parameters as the \simcentral simulation while considering a different distribution of initial Fourier phases. Given that all \lacehc simulations use the same initial Fourier phases, \simseed is useful to evaluate the impact of cosmic variance in the training set on \forestflow predictions.
    
    \item The \simh, \simnu, and \simcurved simulations adopt the same values of $\Delta^2_\mathrm{p}(z=3)$, $n_\mathrm{p}(z=3)$, physical cold dark matter and baryonic densities, and astrophysical parameters as the \simcentral simulation. However, the \simh simulation uses 10\% larger Hubble parameter ($h=0.74$) and 18\% smaller matter density ($\Omega_\mathrm{M}=0.259$) while using the same value of $\Omega_\mathrm{M} h^2$ as the \lacehc simulations, the \simnu simulation includes massive neutrinos ($\sum m_\nu=0.3$ eV), and the \simcurved simulation considers an open universe ($\Omega_k=0.03$). The \simnu and \simcurved simulations also modify the value of the cosmological constant while holding fixed $h$ to compensate for the increase in the matter density and the addition of curvature, respectively. We use the testing simulations to evaluate the performance of the emulation strategy for cosmologies not included in the training set.

    \item The \simigm simulation uses the same cosmological parameters as the \simcentral simulation while implementing a distinct helium ionization history relative to the \simcentral and \lacehc simulations \citep{puchwein2019ConsistentModellingMetagalactic}. The main difference between the ionization histories of these simulations is that the one implemented in the \simigm simulation peaks at a later time than the others, leading to a significantly different thermal history. The \simigm simulation therefore tests the performance of \forestflow for thermal histories not considered in the \lacehc simulations.
\end{itemize}


\subsection{Simulating \lyaf data}
\label{sec:input_extract_lya}

To extract \lyaf measurements from each simulation, we first selected one of the simulation axes as the line of sight and displace the simulation particles from real to redshift space along this axis. Then, we computed the transmitted flux fraction along $768^2$ uniformly distributed line of sights along this axis using \textsc{FSFE}\footnote{\url{https://github.com/sbird/fake_spectra}} \citep{bird2017FSFEFakeSpectra}; these lines of sight are commonly known as skewers. Following \citet{Gadget_Springel}, we computed pressure forces using the density-entropy formulation of smoothed particle hydrodynamics (SPH) with a cubic spline kernel and 33 neighbors\footnote{This approach may lead to biases in flux statistics \citep{hydro_Chabanier2023}.}. We set the resolution of the skewers to 0.05 Mpc, which is enough to resolve the thermal broadening and pressure scales, and spaced these by 0.09 Mpc in the transverse direction. We checked that \pthreed and \poned measurements within the range of interest (see Sect.~\ref{sec:input_fitting}) do not vary by increasing the line-of-sight resolution or the transverse sampling. We repeated all the previous steps for the three simulation axes to extract further cosmological information, as each simulation axis samples the velocity field in a different direction. Finally, we scaled the effective optical depth of the skewers to 0.90, 0.95, 1.05, and 1.10 times its original value, which is equivalent to running simulations with different UV background photoionization rates \citep[see][]{hydro_Lukic2015}.

Using this data as input, we measured \pthreed by first computing the three-dimensional Fourier transform of the skewers. Then, we took the average of the square norm of all modes within 20 logarithmically spaced bins in wavenumber $k$ from the fundamental mode of the box, $k_\mathrm{min}=2\uppi L^{-1} \simeq 0.09\iMpc$, to $k_\mathrm{max}=40\iMpc$ and 16 linearly spaced bins in the cosine of the angle between Fourier modes and the line of sight from $\mu=0$ to 1. We measured \poned by first computing the one-dimensional Fourier transform of each skewer without applying any binning, and then by taking the average of the square norm of all these Fourier transforms. The impact of cosmic variance on fixed-and-paired simulations is not straightforward \citep{maion2022fpvariance}, and thus we would ideally use multiple fixed-and-paired simulations with different initial distributions of Fourier phases to estimate the precision of \pthreed and \poned measurements from our simulations. However, such simulations are not available, and we instead relied on the comparison between two simulations with same configuration but different initial conditions (see Appendix~\ref{sec:cosmic_variance}). We found that the impact of cosmic variance on \pthreed and \poned measurements can be as large as 10 and 1\% on intermediate scales, respectively.

We measured \pthreed and \poned from the 30 \lacehc and the six test simulations, ending up with 2 (opposite Fourier phases) $\times$ 3 (simulation axes) $\times$ 11 (snapshots) $\times$ 5 (mean flux rescalings) $=330$ measurements of each of these statistics per simulation. To reduce cosmic variance, we computed the average of measurements from different axes and phases of fixed-and-paired simulations, which decreased the number of measurements per simulation to 55. The training and testing sets of \forestflow are thus comprised of 1650 and 330 \lya power spectra measurements\footnote{These are publicly available at \url{https://github.com/igmhub/LaCE}}, respectively.


\subsection{Fitting the parametric model}
\label{sec:input_fitting}

To generate training and testing data for our emulator, we computed the best-fitting parameters of Eqs.~\ref{eq:p3d_model} and ~\ref{eq:dnl} to measurements from the simulations described in Sect.~\ref{sec:input_sims}. We fitted the model using \pthreed measurements from $k=0.09$ to $5\iMpc$ and \poned measurements from $k_\parallel=0.09$ to $4\iMpc$. The size of our simulation boxes determines the largest scales used, while the smallest scales measured by DESI set the maximum wavenumbers \citep{ravoux2023DarkEnergySpectroscopica, karacayli2024Optimal1DLy}. It is important to note that the two \lya linear biases determine the large-scale behavior of \pthreed (see Eq.~\ref{eq:p3d_model}), and thus \forestflow could make accurate predictions for \pthreed on arbitrarily large (linear) scales as long as the \lya linear biases are measured accurately.

We computed the best-fitting value of model parameters $\mathbf{y}=\{b_\delta,\, b_\eta,\, q_1,\, q_2,\, k_\mathrm{v},\, a_\mathrm{v},\, b_\mathrm{v}, \, k_\mathrm{p}\}$ to simulation measurements by minimizing the pseudo-$\chi^2$:
\begin{align}
    \label{eq:chi2}
    \chi^2(\mathbf{y}) =& \,\,\sum_{i}^{M_\mathrm{3D}} w_\mathrm{3D}\left[P_\mathrm{3D}^\mathrm{data}(k_i, \mu_i) - P_\mathrm{3D}^\mathrm{model}(k_i, \mu_i, \mathbf{y})\right]^2 \nonumber\\ 
    +&\,\, \sum_{i}^{M_\mathrm{1D}} w_\mathrm{1D} 
    \left[P_\mathrm{1D}^\mathrm{data}(k_{\parallel,\,i}) - P_\mathrm{1D}^\mathrm{model}(k_{\parallel,\,i}, \mathbf{y})\right]^2,
\end{align}
where $M_\mathrm{3D}=164$ and $M_\mathrm{1D}=42$ were the number of \pthreed and \poned bins employed in the fit, respectively, the superscripts data and model refer to simulation measurements and model predictions, and $w_\mathrm{3D}$ and $w_\mathrm{1D}$ weighed the fit. We used the Nelder-Mead algorithm implemented in the routine {\sc minimize} of {\sc scipy} \citep{virtanen2020_SciPyFundamentalalgorithms} to carry out the minimization\footnote{To ensure that this routine did not get stuck in a local minimum, we checked that the likelihood is unimodal in all cases using the Affine Invariant Markov chain Monte Carlo Ensemble sampler \textsc{emcee} \citep{foremanmackey13}.}. The results of the fits are publicly accessible\footnote{\url{https://github.com/igmhub/ForestFlow}}.

Ideally, we would have used the covariance of \pthreed and \poned measurements to weigh the previous expression. However, estimating this covariance requires multiple realizations of the same simulation with different initial distributions of Fourier phases, and we do not have these simulations available. Instead, we disregarded correlations between \pthreed and \poned and weighed these by $w_\mathrm{3D}= N_\mathrm{3D}(k, \mu)/(1+\mu^2)^2$ and $w_\mathrm{1D}=\alpha (1+k_{\parallel}/k_0)^2$, where $N_\mathrm{3D}$ is the number of modes in each $k-\mu$ bin and $k_0=2\iMpc$. The terms involving $N_\mathrm{3D}$, $\mu$, and $k_0$ attempt to ensure an unbiased fit of \pthreed and \poned across the full range of scales used. The parameter $\alpha=8000$ controls the relative weight of \pthreed and \poned in the fit, and we set this value motivated by the different impact of cosmic variance on these statistics (see Appendix~\ref{sec:cosmic_variance}).

We expect significant correlations between the best-fitting value of the parameters to measurements from relatively small simulation boxes. As shown by \citet{arinyo-i-prats2015NonlinearPowerSpectrum}, these correlations are especially significant for the parameters accounting for nonlinear growth of structure, $q_1$ and $q_2$. \citet{givans2022NonlinearitiesLymanAlpha} advocated for setting $q_2=0$ since this parameter is not necessary for describing \pthreed at $z=2.8$. However, we found non-zero values of this parameter indispensable for describing \pthreed at redshifts below $z=2.5$. This is not surprising since the gravitational evolution of density perturbations becomes increasingly more nonlinear as cosmic time progresses.


\begin{figure}
\includegraphics[width=\columnwidth]{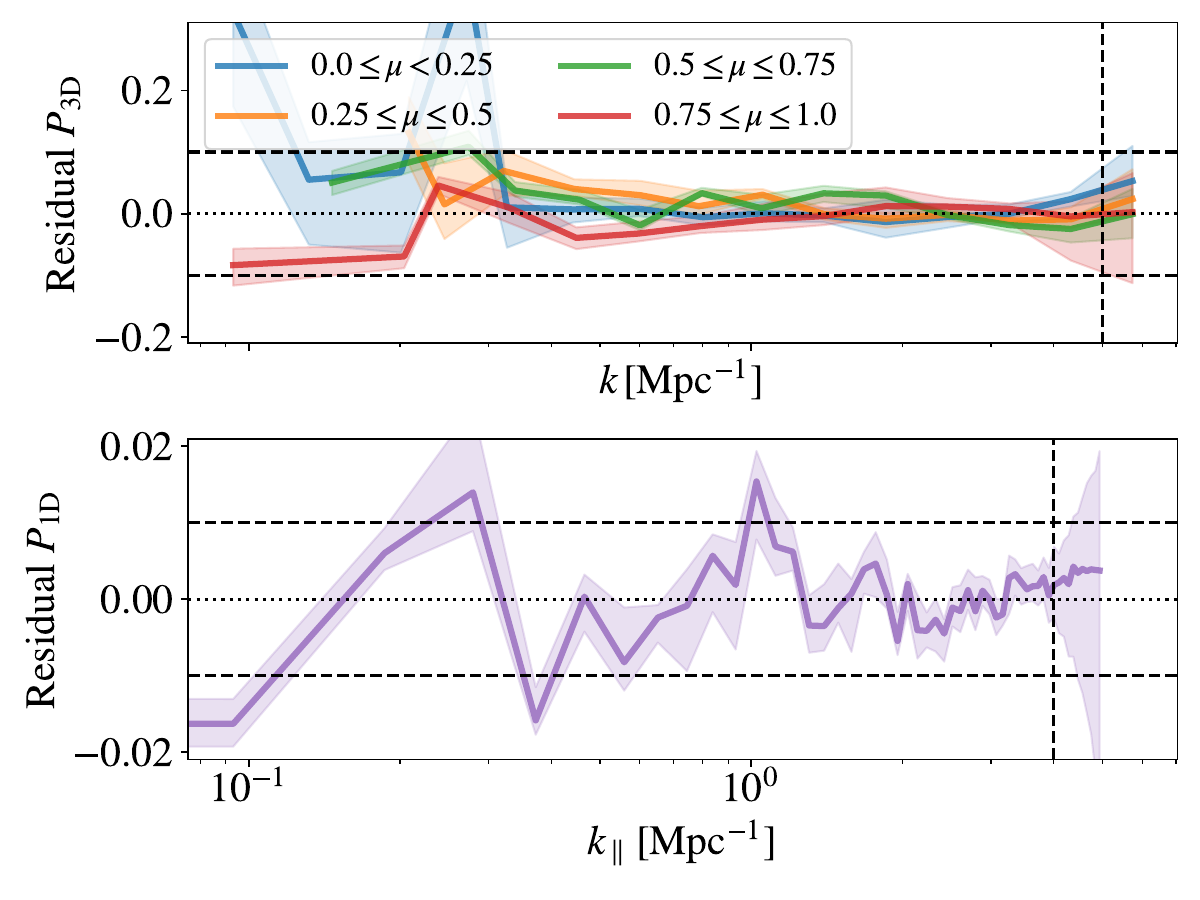}
\centering
\caption{Accuracy of the parametric model (see Eqs.~\ref{eq:p3d_model} and \ref{eq:dnl}) in reproducing \pthreed and \poned measurements from all the \lacehc simulations. Lines and shaded areas show the mean and standard deviation of the relative difference between simulation measurements from the 1650 snapshots of the \lacehc simulations and best-fitting models to these, respectively. The accuracy of the model in recovering \pthreed and \poned is 2.4 and 0.6\%, respectively, on scales not strongly affected by cosmic variance.}
\label{fig:goodness}
\end{figure}

\subsection{Accuracy of the parametric model}
\label{sec:input_precision}

In the previous section, we computed the best-fitting parameters of the \pthreed model to measurements from the \lacehc simulations. Two main sources of uncertainty can affect these fits: model inaccuracies and cosmic variance. The first relates to using a parametric model without enough flexibility to describe \lya clustering accurately, while the second has to do with the limited size of the \lacehc simulations. The influence of cosmic variance on the training set is amplified because all the \lacehc simulations use the same initial distribution of Fourier phases, meaning all simulations are subject to the same large-scale noise. We study this source of uncertainty in Appendix~\ref{sec:cosmic_variance}, where we compared the best-fitting models to the \simcentral and \simseed simulations, whose only difference is in their initial distribution of Fourier phases. We proceed to study model inaccuracies next.

In Fig.~\ref{fig:goodness}, we show the performance of the parametric model in reproducing \pthreed and \poned measurements from the 1650 snapshots of the \lacehc simulations. As discussed in Sect.~\ref{sec:strategy_model}, cosmic variance limits our ability to evaluate the accuracy of the model for \pthreed on scales $k<0.5\iMpc$; therefore, we quote the model accuracy from $k=0.5\iMpc$ down to the smallest scale used in the fit, $k=5\iMpc$. In contrast, since cosmic variance has a much smaller impact on \poned, we evaluate the model performance for this statistic using all scales considered in the fit ($0.09<k_\parallel[\iMpc]<4$). We adopt the same approach when evaluating the performance of \forestflow in Sect.~\ref{sec:results}. Under these considerations, the overall accuracy of the parametric model is 2.4 and 0.6\% for \pthreed and \poned, respectively. Given that we estimate the accuracy of the parametric model using the \lacehc simulations, the previous numbers account for both the limited flexibility of such model and cosmic variance. As discussed in Appendix~\ref{sec:cosmic_variance}, the impact of cosmic variance on measurements of \pthreed and \poned from these simulations is 1.3 and 0.5\%, respectively. If we assume that cosmic variance and errors coming from the limited flexibility of the parametric model are uncorrelated and add in quadrature, the second are responsible for 2.0 and 0.3\% errors on \pthreed and \poned, respectively.


\begin{figure*}
    \includegraphics[width= 0.98\textwidth]{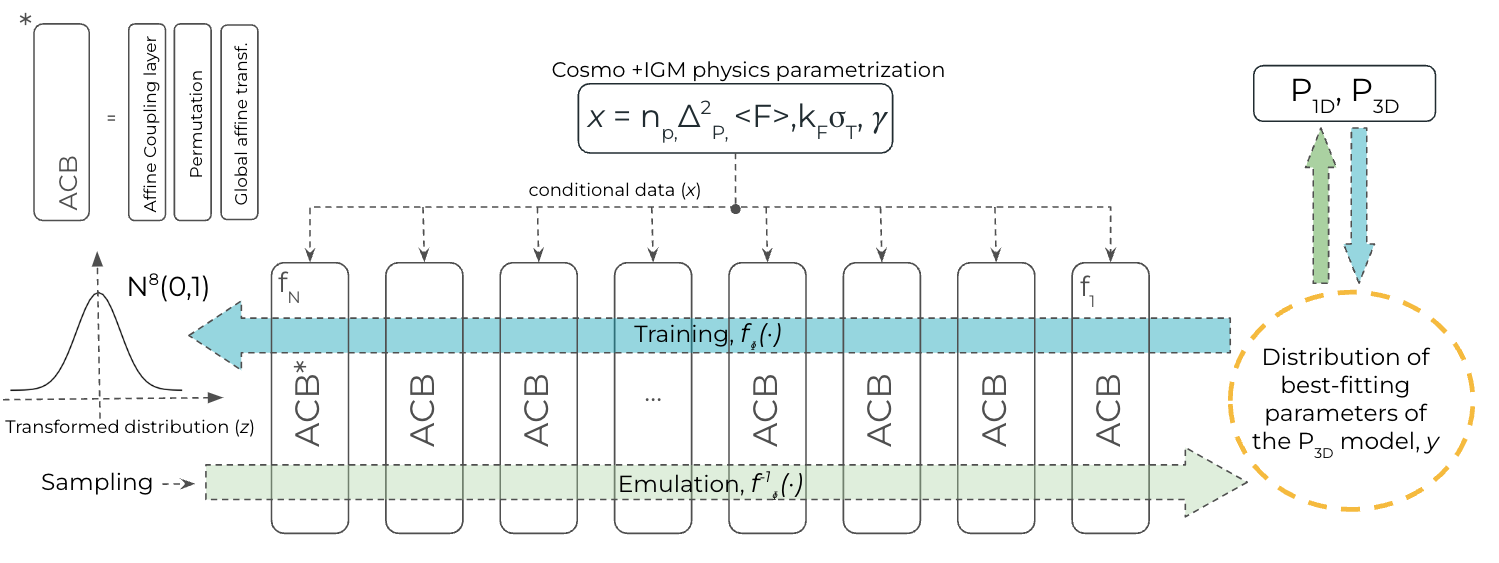} 
    \centering
    \caption{Architecture of the \lyaf clustering emulator. The blue arrow indicates the training direction, where the cNF optimizes a bijective mapping between the best-fitting parameters of the \pthreed model to measurements from the \lacehc simulations and an eight-dimensional Normal distribution. The mapping is conditioned on cosmology and IGM physics, and performed using 12 consecutive affine coupling blocks. The green arrow denotes the emulation direction, where the cNF applies the inverse of the mapping to random samples from the base distribution to predict the value of the \pthreed model parameters. Outside the cNF, \forestflow introduces these parameters in Eq.~\ref{eq:p3d_model} and \ref{eq:p1d} to obtain predictions for \pthreed and \poned, respectively.}
    \label{fig:net_architecture}
 \end{figure*}

\section{Emulator}
\label{sec:forestflow}

In this section, we use a cNF to predict the two \lya linear biases and six parameters describing small-scale deviations of \pthreed from linear theory as a function of cosmology and IGM physics. We detail the architecture and implementation of this emulator in Sect.~\ref{sec:forestflow_NF} and \ref{sec:forestflow_implementation}, respectively.


\subsection{Conditional normalizing flows}
\label{sec:forestflow_NF}

Normalizing flows \citep[NFs;][]{NF_Rezende2015} are a class of machine-learning generative models designed to predict complex distributions by applying a sequence of bijective mappings to simple base distributions. A natural extension to this framework is conditional NFs \citep[cNFs;][]{Winkler2019, cNF_Papamakarios}, a type of NFs that condition the mapping between the base and target distributions on a series of input variables. Given an input $\mathbf{x} \in X$ and target $\mathbf{y} \in Y$, cNFs predict the conditional distribution $p_{Y|X}(\mathbf{y}|\mathbf{x})$ by applying a parametric, bijective mapping $f_\phi: Y\times X \to Z$ to a base distribution $p_{Z}(\mathbf{z})$ as follows
\begin{equation}
    \label{eq:cNF_pdf}
    p_{Y|X}(\mathbf{y}|\mathbf{x}) = p_{Z}(f_\phi(\mathbf{y}, \mathbf{x})|\mathbf{x}) \left|\frac{\partial f_\phi(\mathbf{y}, \mathbf{x})}{\partial \mathbf{y}}\right|,
\end{equation}
where $\phi$ are the parameters of the mapping, while the last term of the previous equation is the Jacobian determinant of the mapping. In our cNF, the input is given by the parameters capturing the dependence of the \lyaf on cosmology and IGM physics, $\mathbf{x}=\{\Delta_\mathrm{p}^2,\, n_\mathrm{p},\, \mflux,\, \sigma_\mathrm{T},\, \gamma,\, k_\mathrm{F}\}$, the target by the parameters of the \pthreed model, $\mathbf{y}=\{b_\delta,\, b_\eta,\, q_1,\, q_2,\, k_\mathrm{v},\, a_\mathrm{v},\, b_\mathrm{v}, \, k_\mathrm{p}\}$, and the base distribution is an eight-dimensional Normal distribution $N^8(0,1)$, where the dimension is determined by the number of \pthreed model parameters.

Once trained, cNFs are a generative process from $\mathbf{x}$ to $\mathbf{y}$. In our implementation, we start by randomly sampling from the base distribution, and then we pass this realization through a sequence of mappings conditioned on a particular combination of cosmology and IGM parameters, $\mathbf{\tilde{y}}=f_\phi^{-1}(p_{Z}(\mathbf{z}), \mathbf{x})$, ending up with a prediction for the value of the \pthreed model parameters. Repeating this process multiple times, the emulator yields a distribution of \pthreed parameters $p_{\tilde{Y}|X}$ that, for a sufficiently large number of samples, approaches the target distribution $p_{Y|X}$. The breadth of this distribution captures uncertainties arising from the limited size of the training set. Finally, outside the cNF, we use each combination of \pthreed parameters to evaluate Eqs.~\ref{eq:p3d_model} and \ref{eq:p1d}, obtaining predictions and uncertainties for \pthreed and \poned.

The main challenge when using cNFs is finding the mapping between the target and the base distribution, typically done using an $N$-layer neural network with bijective layers. This process runs in reverse relative to the generating process: we start by applying the mapping $f_\phi$ to the target data $\mathbf{y}$ conditioned on the input $\mathbf{x}$, yielding $\mathbf{z}$. Then, we optimize the model parameters by minimizing the loss function
\begin{equation}
    \mathcal{L} = \frac{1}{2} \sum \textbf{z}^2 - \log \left|\frac{\delta f_\phi(\mathbf{y}, \mathbf{x})}{\delta \mathbf{y}}\right|\,.
    \label{eq:loss} 
\end{equation}
We carried out this optimization process using stochastic gradient descent applied to minibatches, a methodology commonly employed for training neural networks.


\subsection{Implementation}
\label{sec:forestflow_implementation}

Neural Autoregressive Flows \citep{NAF} use a series of invertible univariate operations to build a bijective transformation between a conditional distribution and a base distribution. In \forestflow, we created a bijective mapping between the best-fitting parameters of the \pthreed model and an eight-dimensional Normal distribution by applying $N_\mathrm{ACB}=12$ consecutive Affine-Coupling Block \citep[ACB;][]{RealNVP} conditioned on cosmology and IGM physics. The transformation goes from the best-fitting parameters of the \pthreed model to the base distribution when training the model, and in the opposite direction when evaluating it.

Each ACB conducts a series of operations $g_{i,\tilde{\phi}_i}$ on its input data $\mathbf{w}_i$, with $i$ going from 1 to $N_\mathrm{ACB}$ and $\tilde{\phi}_i$ standing for the parameters of the transformation. First, it splits the input data into two subsamples with approximately the same number of elements, $\mathbf{w'}_i$ and $\mathbf{w''}_i$. Then, it applies an affine transformation to the first subsample $\mathbf{w'}_i$
\begin{equation}
    T(\mathbf{w'}_i)=\alpha_i\, \mathbf{w'}_i + \beta_i,
\end{equation}
where $\alpha_i$ and $\beta_i$ are neural networks with a single hidden layer of 128 neuron units. Third, the ACB merges the output from the affine transformation and the unchanged subsample, and then it applies a permutation layer to randomly rearrange these elements, obtaining $\mathbf{\tilde{w}}_i$. Fourth, the ACB applies an affine transformation to this sample, $\tilde{T}(\mathbf{\tilde{w}}_i)$. The first and second affine transformations involve a subset of the training set and the entire training set, respectively, enabling the model to capture local and global features.

In Fig.~\ref{fig:net_architecture}, we show the architecture of the cNF. The blue arrow indicates the training direction, while the green arrow depicts the emulation direction. In the training direction, the input to the first ACB, $\mathbf{u}_1=\mathbf{w}_1$, is a 1650-dimensional array composed of 14-dimensional vectors, where 1650 is the number of simulation snapshots in the training set. Each vector includes the eight best-fitting \pthreed model parameters to each snapshot and the six parameters describing the cosmology and IGM physics of this snapshot. The input to the $i$ ACB, $\mathbf{u}_i$, is a 1650-array containing 14-dimensional vectors with the output of the $i-1$ ACB and, once again, the six parameters describing the cosmology and IGM physics of each snapshot. Each ACB applies a transformation $f_{i,\phi_i}=g_{i,\tilde{\phi}_i}$, and the consecutive application of all ACBs results in the mapping between the target and the base distributions $\mathbf{z}=f_\phi(\mathbf{y}, \mathbf{x})$, where $f_\phi = \prod_{i=1}^{N_\mathrm{ACB}} f_{i,\phi_i}$.

In the emulation direction, the input to the first ACB, $\mathbf{v}_1=\mathbf{w}_1$, is a 14-dimensional vector containing random draws from an eight-dimensional Normal distribution and the six parameters describing the cosmology and IGM physics for which we want to obtain predictions. As in the training direction, the input to each subsequent ACB relies on the output from the previous ACB, each conditioned on cosmology and IGM physics. The ACBs apply the transformations $f^{-1}_{i,\phi_i}=g_{i,\tilde{\phi}_i}$, which are the inverse of the corresponding transformations in the training direction, $f_{i,\phi_i}$. The cNF makes predictions for \pthreed model parameters by applying the composition of the inverse of all ACBs to random samples from the base distribution, $\mathbf{\tilde{y}}=f_\phi^{-1}(p_{Z}(\mathbf{z}), \mathbf{x})$, where $f_\phi^{-1} = \prod_{i=1}^{N_\mathrm{ACB}} f_{i,\phi_i}^{-1}$.

We implemented the emulator within the \texttt{FreIA} framework \citep{freia}, which uses \texttt{PyTorch} \citep{Ansel_PyTorch_2_Faster_2024} in the backend. We trained it by minimizing Eq.~\ref{eq:loss} using an \texttt{Adam} optimizer \citep{adam_Diederik2015} for 300 epochs with an initial learning rate of $10^{-3}$. We used the \texttt{Optuna} framework \citep{optuna_2019} to select the number of ACBs and epochs, as well as the value of the learning rate. First, \texttt{Optuna} trains our cNF for a particular combination of these hyperparameters. Then, it computes the average value of Eq.~\ref{eq:chi2} for all simulations in the training set. After that, depending on the goodness of the fit to \pthreed and \poned measurements, \texttt{Optuna} selects a new value of the hyperparameters. We iterated with \texttt{Optuna} 50 times through a hyperparameter grid, selecting the hyperparameters that yield the highest accuracy. We checked that the performance of the cNF depended weakly on small variations in the value of the hyperparameters.


\section{Performance of \forestflow}
\label{sec:results}

In Sect.~\ref{sec:results_statistics}, we analyze the performance of \forestflow across the parameter space of the training set. Then, in Sect.~\ref{sec:results_other}, we test its accuracy using simulations with cosmologies and IGM models that are not part of the training set. All performance evaluations are conducted using the simulations described in Sect.~\ref{sec:input_sims}, which were run using the same code and resolution. Before employing \forestflow for cosmological inference, it will be crucial to validate it against large, high-resolution simulations produced with alternative codes. We defer this task to future work.


\subsection{Cosmologies and IGM histories in the training set}
\label{sec:results_statistics}

\begin{figure}
\includegraphics[width= 0.95\columnwidth]{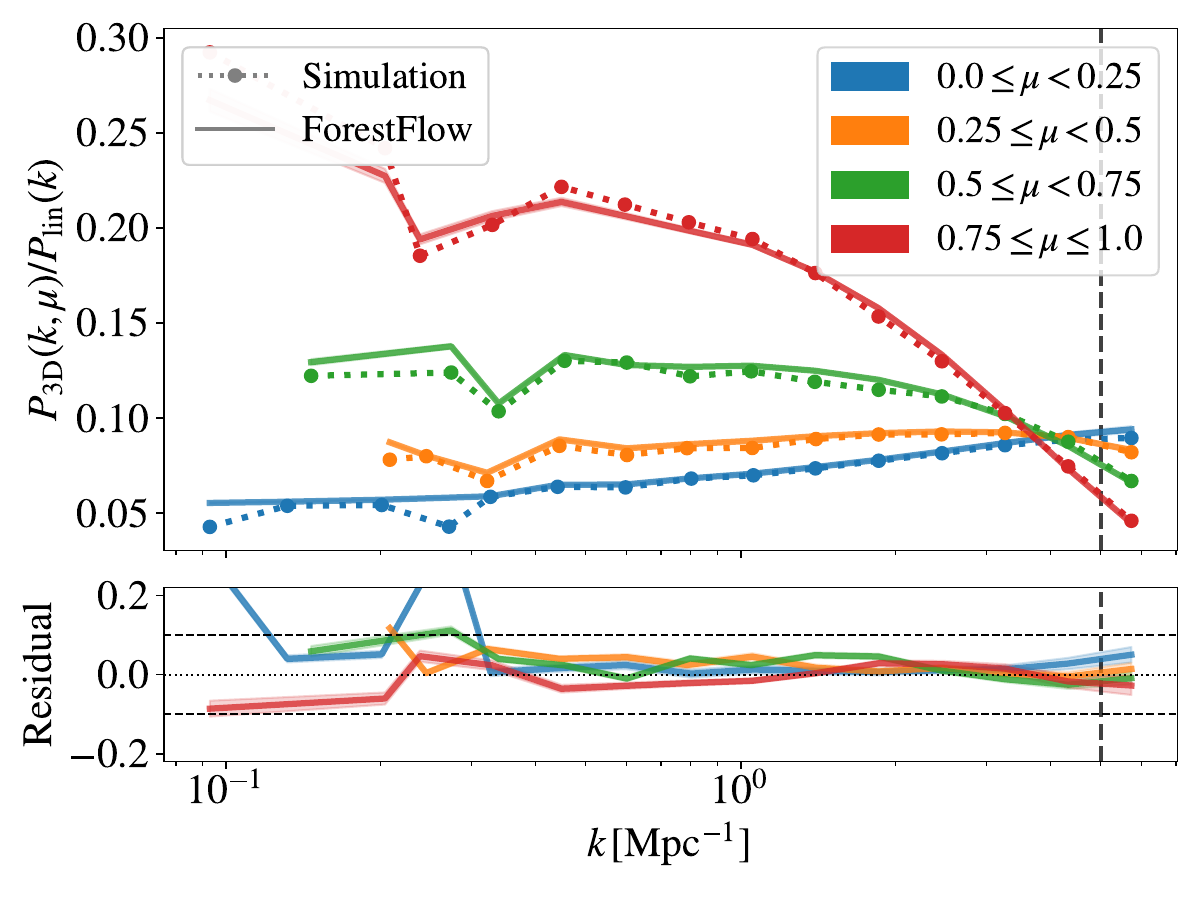}
\includegraphics[width= 0.97\columnwidth]{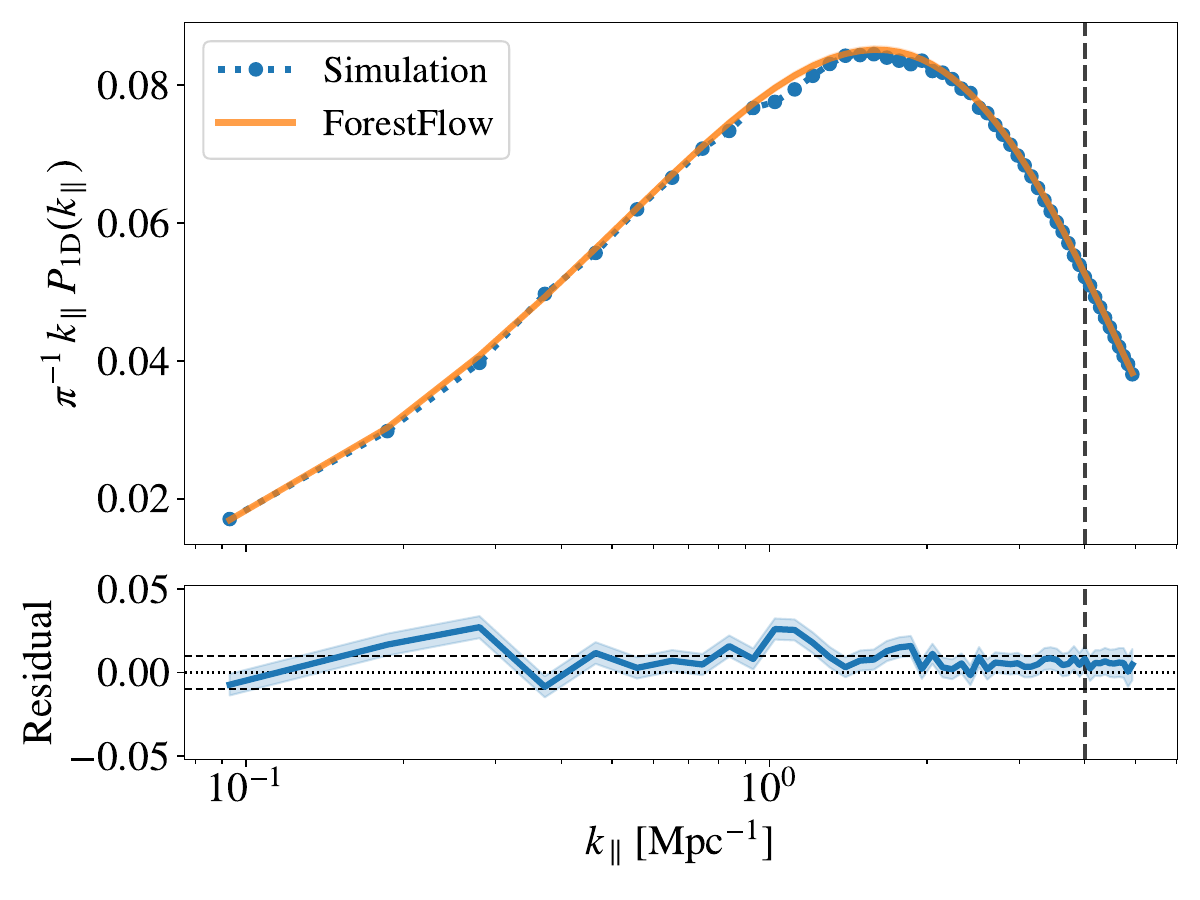}
\centering
\caption{Accuracy of \forestflow in recovering \pthreed and \poned measurements from the \simcentral simulation at $z=3$. Dotted lines show measurements from simulations, solid lines and shaded areas display the average and 68\% credible interval of \forestflow predictions, respectively, and vertical dashed lines indicate the minimum scales considered for computing the training data for the cNF. The overall performance of \forestflow in recovering \pthreed is 2.0\% on scales not strongly affected by cosmic variance and 0.6\% for \poned.
}
\label{fig:test_snap}
\end{figure}

\begin{figure}
\includegraphics[width= 0.95\columnwidth]{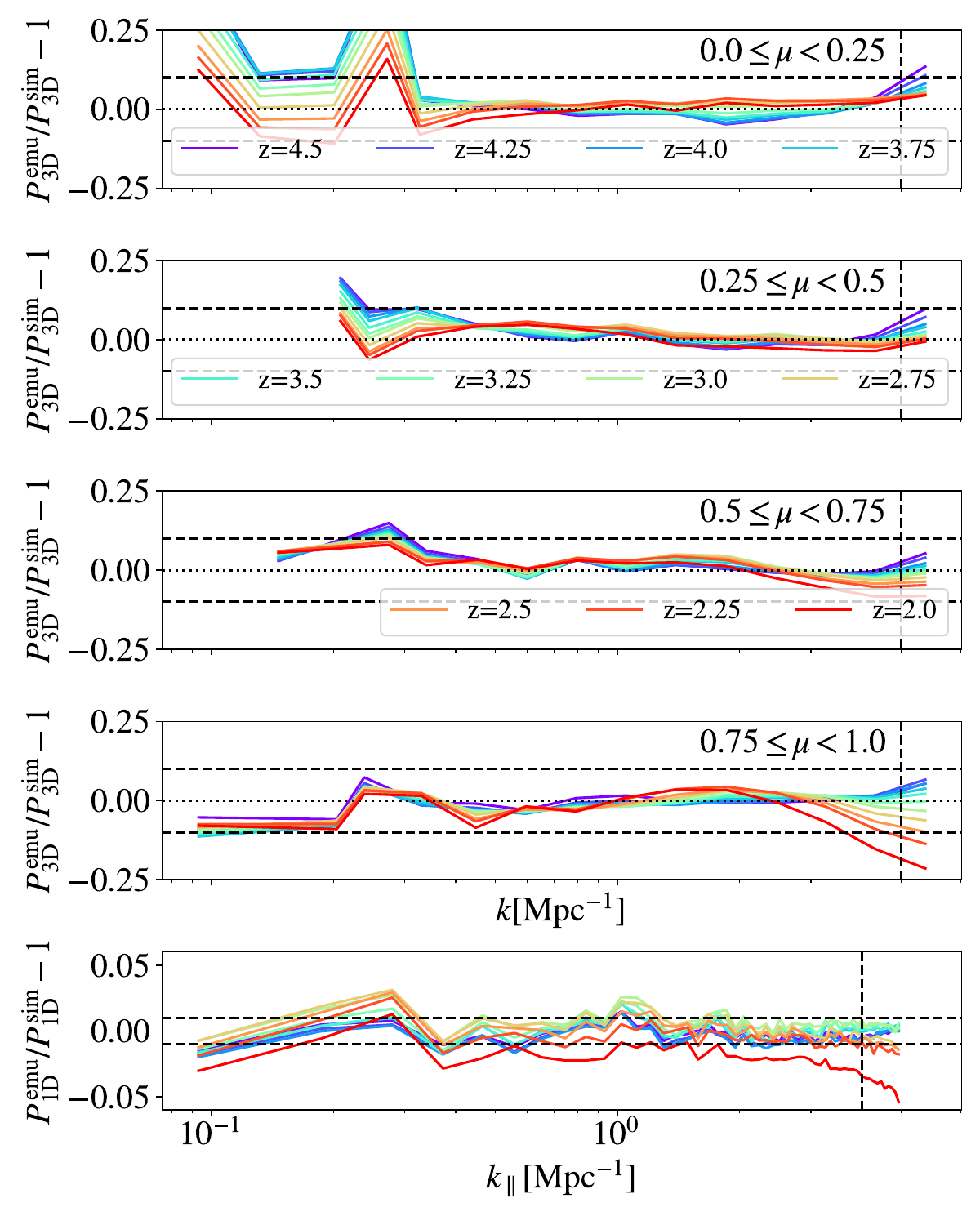}
\centering
\caption{Accuracy of \forestflow in recovering \pthreed and \poned measurements from the \simcentral simulation as a function of redshift. The upper four panels show the results for \pthreed across different $\mu$ bins, while the bottom panel displays the results for \poned. Each color represents a different redshift. The model's accuracy remains consistent for redshifts above $z=2$ but exhibits a slight decline at this redshift.
}
\label{fig:central_z}
\end{figure}

\begin{figure*}
\centering
\Large{Leave-simulation-out}\par\medskip
\includegraphics[width= 0.95\columnwidth]{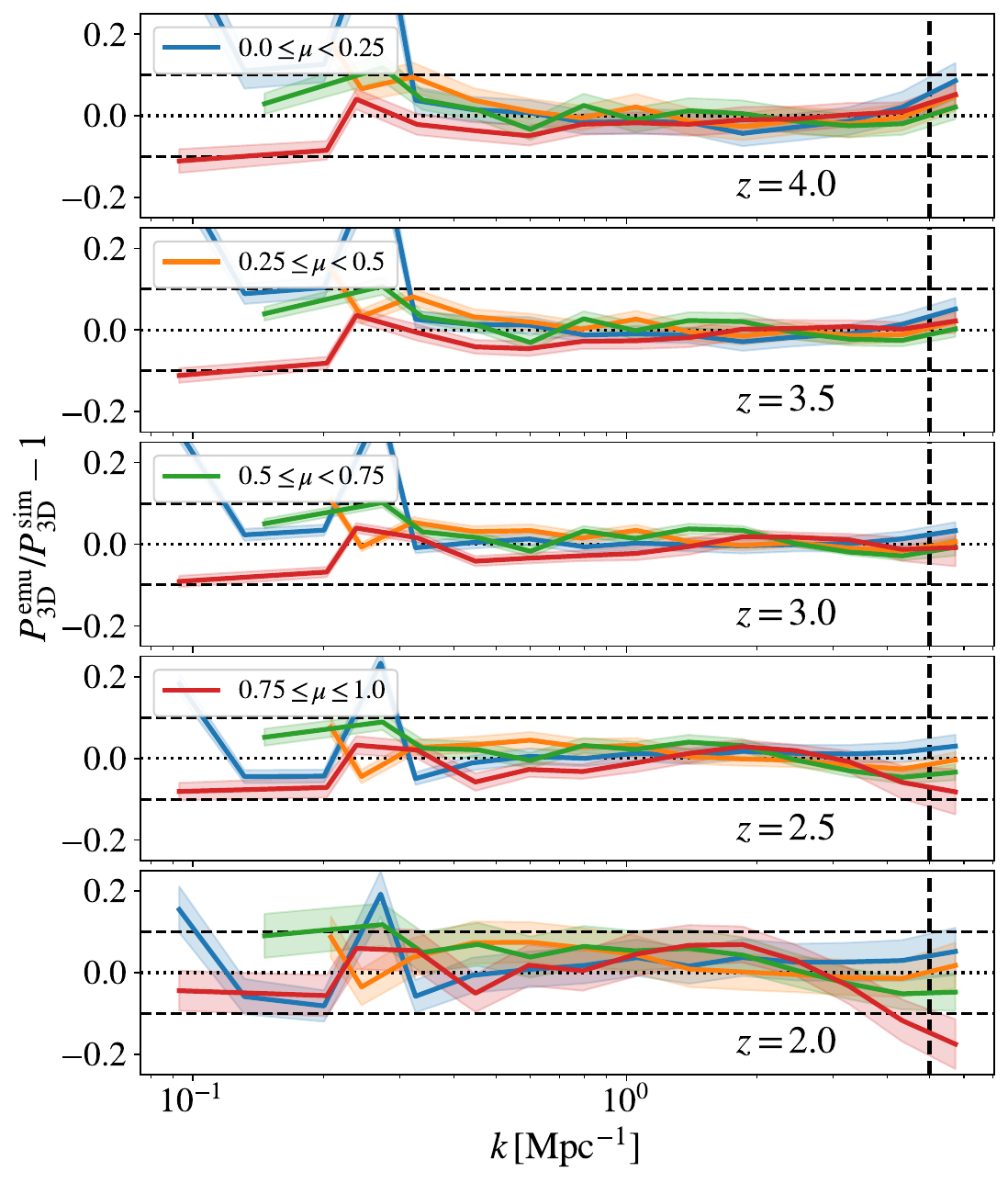}
\includegraphics[width= 0.98\columnwidth]{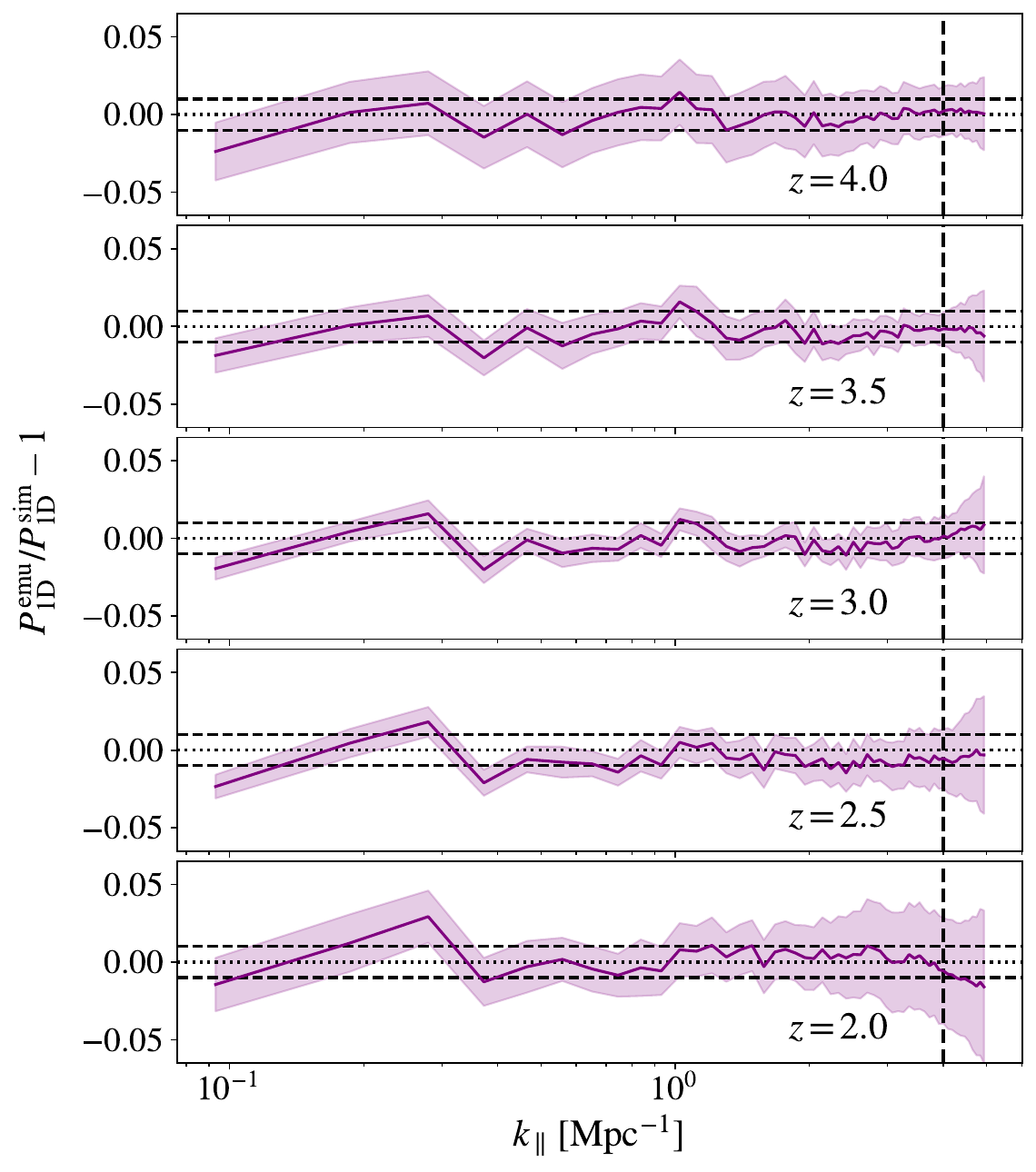}
\Large{Leave-redshift-out}\par\medskip
\includegraphics[width= 0.95\columnwidth]{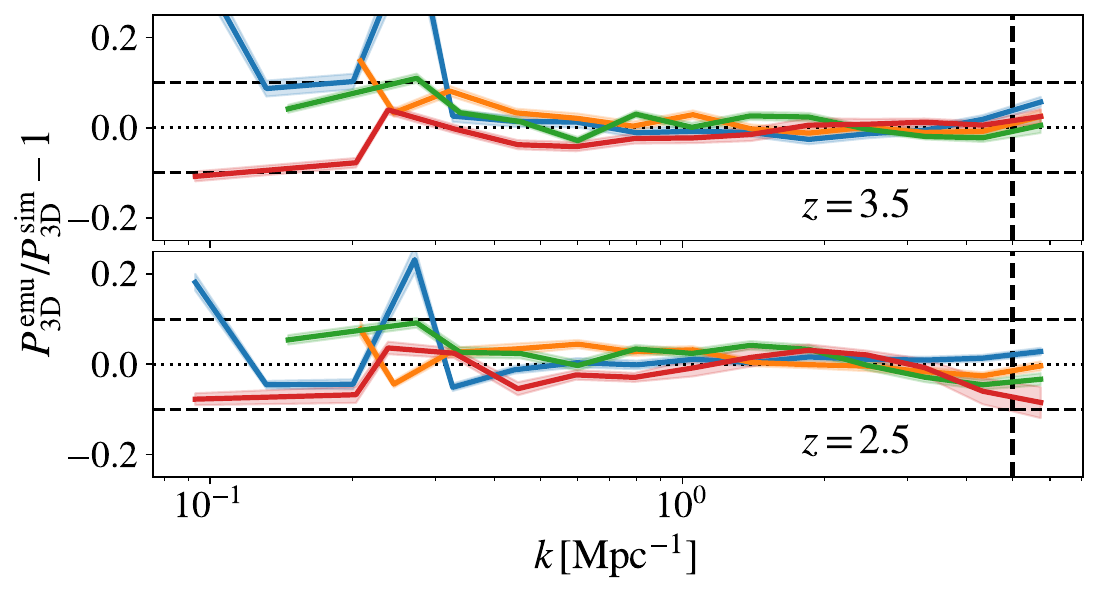}
\includegraphics[width= 0.96\columnwidth]{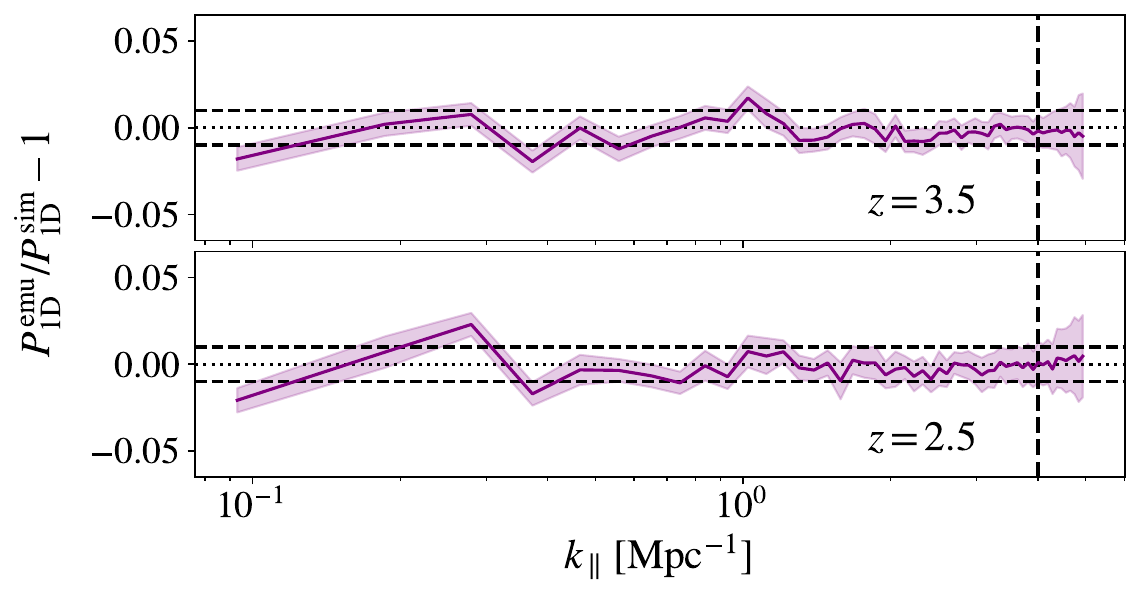}
\caption{Accuracy of \forestflow across the input parameter space estimated via leave-simulation-out (top panels) and leave-redshift-out tests (bottom panels). {\bf Top panels.} Each leave-simulation-out test involves training one independent emulator with measurements from 29 distinct simulations, and then using the measurements from the remaining simulation as the validation set. Lines and shaded areas show the average and standard deviation of 30 leave-simulation-out tests, and each panel shows the results for a different redshift. {\bf Bottom panels.} Leave-redshift-out tests require optimizing one emulator with all measurements but the ones at a particular redshift, and then using measurements from this redshift as validation. Each panel shows the results of a different test.
}
\label{fig:leave_sim_out}
\end{figure*}

In this section, we evaluate the performance of \forestflow in recovering the two Lyman-$\alpha$ linear biases, which determine the behavior of \pthreed on linear scales, as well as \pthreed and \poned measurements from simulations on the intervals $0.5<k[\iMpc]<5$ and $0.09<k_\parallel[\iMpc]<4$, respectively. These are the ranges of scales used when fitting the parametric model in Sect.~\ref{sec:input} that are not strongly affected by cosmic variance (see Sect.~\ref{sec:strategy_model}). We begin by assessing the accuracy of \forestflow at the center of the training set, where machine-learning methods typically perform best, and then extend our evaluation across the entire input parameter space.

In Fig.~\ref{fig:test_snap}, we compare measurements of \pthreed and \poned from the \simcentral simulation at $z=3$ with \forestflow predictions. Dotted lines show simulation measurements, while solid lines and shaded areas display the average and 68\% credible interval of \forestflow predictions, respectively. We characterize the accuracy of the credible intervals in Appendix~\ref{sec:uncertainty_validation}. As we can see, \forestflow captures the amplitude and scale-dependence of \pthreed and \poned precisely. 
In Fig.~\ref{fig:central_z}, we present the relative difference between \pthreed and \poned measurements from the \simcentral simulation and \forestflow predictions as a function of redshift. The model's accuracy remains consistent for redshifts above $z=2$ but shows a slight decline at this redshift. This is likely because $z=2$ is the lowest redshift included in the training set and is therefore near the boundary of the convex hull defined by the training data.

To better characterize the performance of \forestflow, we compute the average accuracy of \forestflow in recovering measurements from \simcentral across redshift. We find that it is 1.2 and 0.3\% for $b_\delta$ and $b_\eta$, respectively, which translates into 1.1 and 1.2\% for perpendicular and parallel \pthreed modes on linear scales, and 2.6 and 0.8\% for \pthreed and \poned. Note that cosmic variance hinders our ability to test the performance of the model; however, this does not necessarily indicate a decrease in the model's accuracy for \pthreed on the largest scales sampled by our simulation.

We expect the efficiency of \forestflow to decrease away from the center of the input space. We could assess its performance across the parameter space using the training simulations; however, the cNF has been optimized for these points, which introduces the risk of overfitting. Overfitting could result in high precision for these specific points of the parameter space but not for others nearby. As a result, we would ideally evaluate the performance of \forestflow using multiple test simulations covering the entire parameter space, but such simulations are unavailable. Instead, we conduct leave-one-out tests, which are widely used to assess the performance of an emulator when the number of training points is insufficient for out-of-sample tests \citep[e.g.,][]{hastie01statisticallearning}. In a leave-one-out test, we optimize a cNF after removing a subsample from the training set; for example, all measurements from one of the \lacehc simulations. We then check the accuracy of \forestflow for the new cNF using the subsample held back. The rationale is that the new emulator should closely approximate the original emulator everywhere in the parameter space except near the excluded simulation, and more importantly, there is no risk of overfitting. By repeating this process for other subsamples, we can estimate the performance of \forestflow across the parameter space. Since each cNF is trained without using the entire dataset, leave-one-out tests provide a lower bound on \forestflow performance. Additionally, leave-one-out tests may require extrapolating the predictions from the cNF, and it is widely known that machine-learning methods do not extrapolate well.

In the top panels of Fig.~\ref{fig:leave_sim_out}, lines and shaded areas display the average and standard deviation of 30 leave-simulation-out tests. Each test requires optimizing a cNF with 29 distinct \lacehc simulations, and then using the remaining simulation as the validation. Each panel shows the results for a different redshift, and we check that the results are similar for redshifts not shown. As we can see, the large-scale noise is similar for all \lacehc simulations; this is because they use the same initial distribution of Fourier phases. The overall performance of \forestflow in recovering $b_\delta$ and $b_\eta$ is 1.0 and 3.1\%, respectively, which translates into 2.0 and 2.9\% for perpendicular and parallel \pthreed modes on linear scales, and 3.4 and 1.8\% for \pthreed and \poned.

\begin{table}
    \caption{
        Percentage impact of different sources of uncertainty on \forestflow predictions and overall accuracy.
    }
    \centering
    \begin{tabular}{c|cc|cc|cc}
         Type & \pthreed & \poned & $b_\delta$ & $b_\eta$ & $P_\mathrm{3D, \perp}^\mathrm{lin}$ & $P_\mathrm{3D, \parallel}^\mathrm{lin}$\\ \hline
         C.~var. fit\tablefootmark{a}   & 0.8 & 0.1 & 0.6 & 1.8 & 1.2 & 1.8 \\
         C.~var. data\tablefootmark{b}  & 1.3 & 0.5 &  -- &  -- &   -- &  -- \\
         C.~var. \& fit\tablefootmark{c} & 2.4 & 0.6 &  -- &   -- &   -- &   -- \\
         Emu. center\tablefootmark{d}  & 2.6 & 0.8 & 1.2 & 0.3 & 1.1 & 1.2 \\
         \rowcolor{gray!30}
         Emu. overall\tablefootmark{e} & 3.4 & 1.8 & 1.0 & 3.1 & 2.0 & 2.9 \\
    \end{tabular}
    \tablefoot{
        The second and third columns show the results for \pthreed and \poned over the intervals $0.5<k[\iMpc]<5$ and $0.09<k_\parallel[\iMpc]<4$ intervals, respectively, while the last two columns do so for perpendicular and parallel \pthreed linear modes.
        \tablefoottext{a}{Impact of cosmic variance on the best-fitting \pthreed model to simulation measurements (see Appendix~\ref{sec:cosmic_variance}).}
        \tablefoottext{b}{Impact of cosmic variance on simulation measurements (see Appendix~\ref{sec:cosmic_variance}).}
        \tablefoottext{c}{Joint impact of cosmic variance and the limited flexibility of the \pthreed model (see Sect.~\ref{sec:input_precision}).}
        \tablefoottext{d}{Performance of \forestflow at the center of the parameter space, estimated using the \simcentral simulation (see Sect.~\ref{sec:results_statistics}).}
        \tablefoottext{e}{Accuracy of \forestflow across the parameter space, estimated via leave-simulation-out tests (see Sect.~\ref{sec:results_statistics}).}
    }
    \label{tab:precision}
\end{table}

In Table~\ref{tab:precision}, we gather the accuracy of \forestflow at the center and across the parameter space, as well as the expected level of uncertainties due to cosmic variance and the limited flexibility of the \pthreed model. Due to the limited size of our simulations, the maximum levels of accuracy we can test for \pthreed and \poned are 1.3 and 0.5\% (see Appendix~\ref{sec:cosmic_variance}), respectively. These levels would decrease by evaluating the accuracy of \forestflow using bigger simulations with the same resolution. On the other hand, the combined impact of impact of cosmic variance on the training data and the limited flexibility of the \pthreed model are 2.4 and 0.6\% for \pthreed and \poned, respectively, which is 1.1 and 0.1\% worse than the minimum accuracy we can test for these statistics. At the center of the parameter space, the accuracy of \forestflow for \pthreed and \poned is only 0.2\% worse than the previous levels, letting us conclude that the primary factors limiting the performance of \forestflow at the center of the parameter space are the size of the training simulations and model inaccuracies. 

The efficiency of \forestflow across the parameter space is 1.2 and 1.0\% worse than at the center for \pthreed and \poned, respectively. Consequently, its accuracy would likely improve by increasing the number of training simulations. However, leave-one-out tests significantly underestimate the performance of an emulator at the edges of the training set, especially for a small number of simulations, because it often requires extrapolating the emulator's predictions. We can thus conclude that the quality of the training data, the accuracy of the model, and the number of training simulations have a similar impact on the performance of \forestflow. Given that leave-simulation-out tests tend to provide a lower performance bound, we conclude that the overall accuracy of \forestflow in predicting \pthreed from linear scales to $k=5\iMpc$ is approximately 3\%, and $\simeq1.5\%$ for \poned down to $k_\parallel=4\iMpc$.

As discussed in Sect.~\ref{sec:strategy_params}, \forestflow does not use as input "traditional" cosmological parameters such as $\Omega_m$, $A_s$, or $H_0$. Instead, it uses a set of parameters measured from the outputs of individual simulation snapshots. This strategy enables training \forestflow without specifying the input redshift and making predictions for redshifts not present in the training set. To test this assumption, we carry out two leave-redshift-out tests. The first involves optimizing one emulator with all \lacehc measurements but the ones at $z=2.5$, and then validating it with data from this redshift. For the second, we follow the same approach but using measurements at $z=3.5$. We display the results of these tests in the bottom panels of Fig.~\ref{fig:leave_sim_out}. The performance of \forestflow is similar for leave-redshift-out and leave-simulation-out tests, validating the approach mentioned above. We find similar results for leave-redshift-out tests at other redshifts.


\begin{figure*}
\includegraphics[width= 0.95\columnwidth]{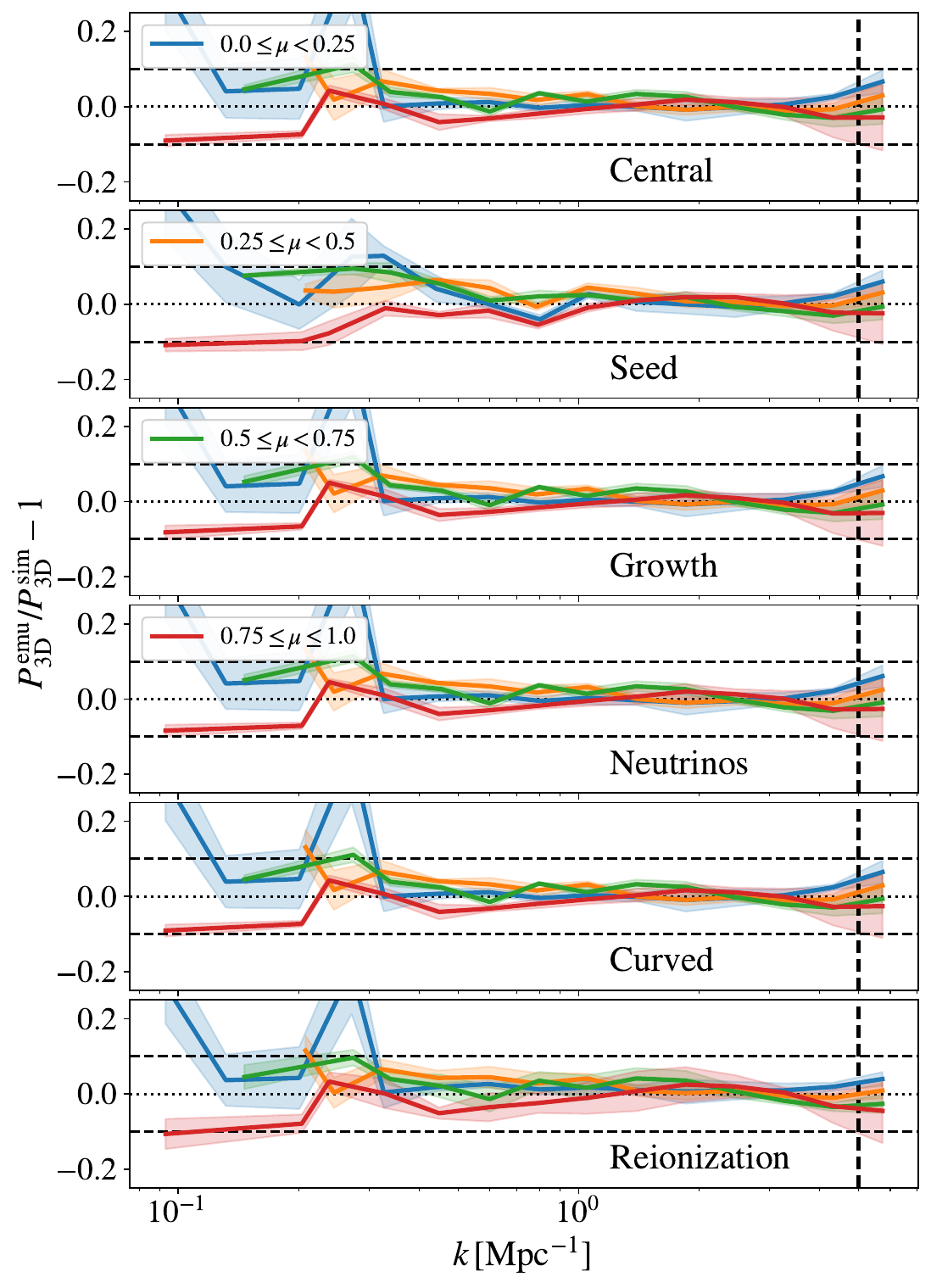}
\includegraphics[width= 0.96\columnwidth]{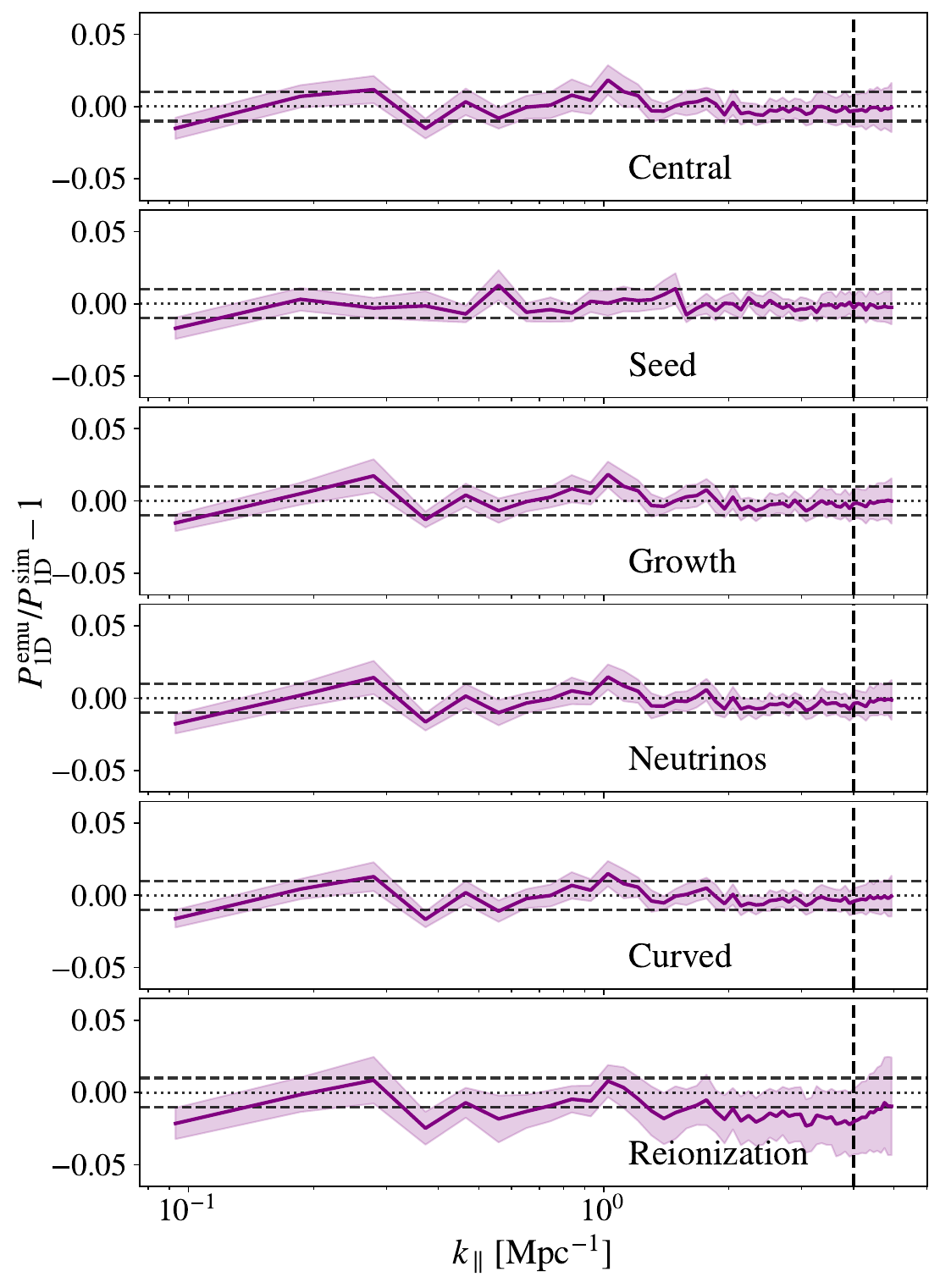}
\centering
\caption{Performance of \forestflow in recovering \pthreed and \poned for test simulations not included in the training set. Lines and shaded areas display the average and standard deviation of the results for 11 snapshots between $z=2$ and 4.5, respectively. From top to bottom, the rows show the results for the \simcentral, \simseed, \simh, \simnu, \simcurved, and \simigm simulations, where the \simcentral and \simseed simulations are at the center of the input parameter space and employ the same and different initial distribution of Fourier phases as the training simulations, respectively, the \simh and \simigm simulations use a different growth and reionization history relative to those used by the \lacehc simulations, and the \simnu and \simcurved simulations consider massive neutrinos and curvature. The efficiency of \forestflow is approximately the same for all simulations.}
\label{fig:other_cosmo}
\end{figure*}

\subsection{Other cosmologies and IGM histories}
\label{sec:results_other}

In Fig.~\ref{fig:other_cosmo}, we examine the accuracy of \forestflow in reproducing \pthreed and \poned measurements from simulations not included in the training set. Lines indicate the redshift average of the relative difference between model predictions and simulation measurements. The first two rows show the results for the \simcentral and \simseed simulations, whose only difference is their initial distribution of phases. Consequently, the predictions of \forestflow are the same for the two simulations. As we can see, these simulations present a different large-scale pattern of fluctuations, signaling that are caused by cosmic variance. Once we ignore these, we find that the performance of \forestflow is practically the same for the two simulations. We can thus conclude that \forestflow predictions are largely insensitive to the impact of cosmic variance on the training set.

In the third, fourth, and fifth rows of Fig.~\ref{fig:other_cosmo}, we use the \simh, \simnu, and \simcurved simulations to evaluate the accuracy of \forestflow for three different scenarios not contemplated in the training set: different growth history, massive neutrinos, and curvature. As we can see, the performance of \forestflow for all these simulations is approximately the same as for the \simcentral simulation. These results support that using the small-scale amplitude and slope of the linear power spectrum to capture cosmological information enables setting precise constraints on growth histories and $\Lambda$CDM extensions not included in the training set \citep[see also][]{Pedersen2021, pedersen2023CompressingCosmologicalInformation, cabayol-garcia2023NeuralNetworkEmulator}.

In the last row of Fig.~\ref{fig:other_cosmo}, we examine the accuracy of \forestflow for the \simigm simulation, which employs a \ion{He}{ii} reionization history significantly different from those used by the \lacehc simulations. The performance of \forestflow for this and the \simcentral simulation is similar, which is noteworthy given that the performance of \poned emulators for the \simigm simulation is significantly worse than for the \simcentral simulation \citep{cabayol-garcia2023NeuralNetworkEmulator}. The outstanding performance of \forestflow is likely because the relationship between IGM physics and the parameters of the \pthreed model is more straightforward than with \poned variations.


\section{Discussion}
\label{sec:discussion}

Cosmological analyses of the \lya forest come in two flavors: one-dimensional studies focused on small, nonlinear scales and three-dimensional analyses of large, linear scales. With \forestflow, we can now consistently model \lya correlations from nonlinear to linear scales, enabling a variety of promising analyses that we discuss next.


\subsection{Connecting large-scale biases with small-scale physics}
\label{sec:discussion_large_small}

\begin{figure*}
\includegraphics[width=\columnwidth]{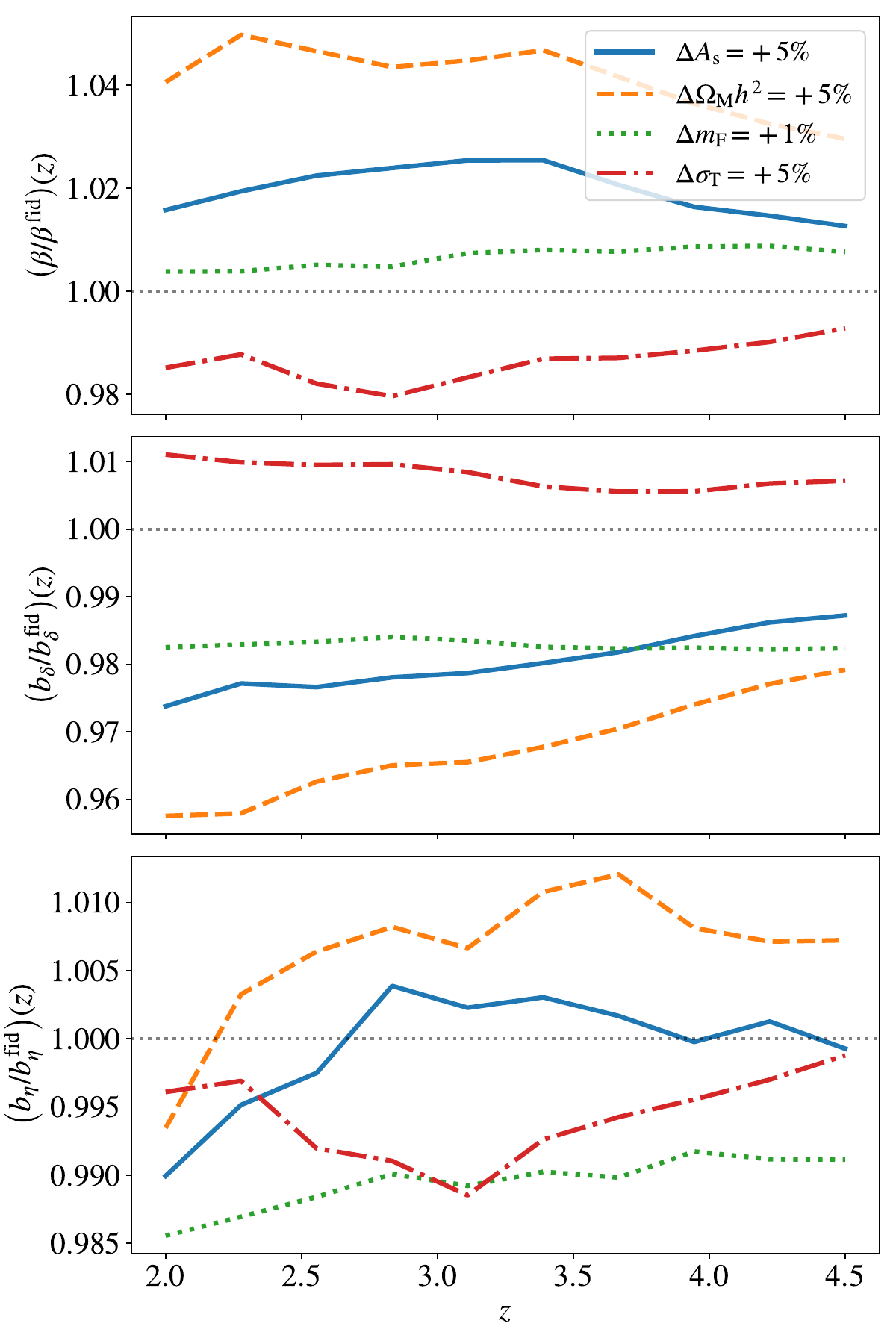}
\includegraphics[width=\columnwidth]{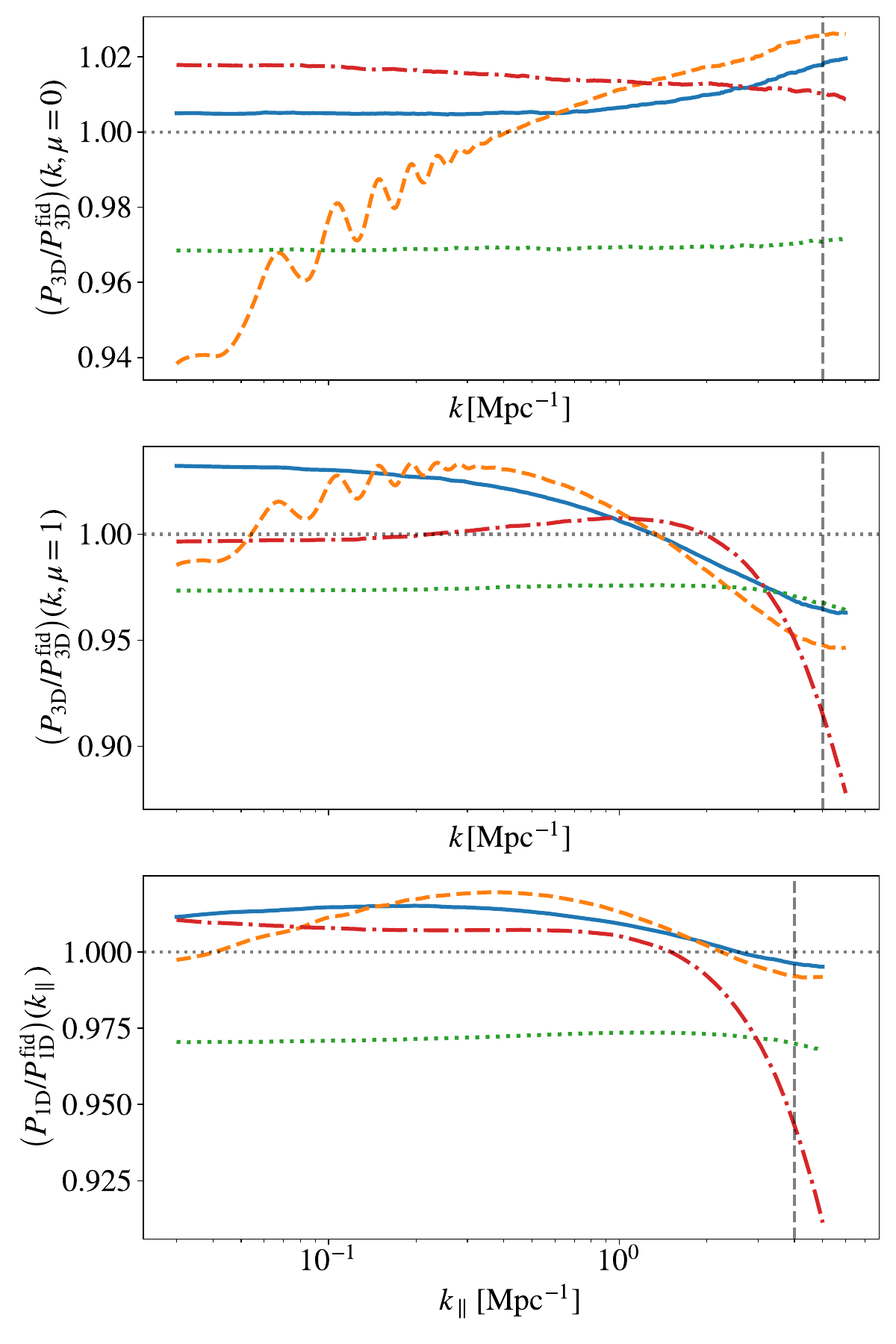}
\centering
\caption{Response of \lya clustering to variations in cosmology and IGM physics according to \forestflow. The top, middle, and bottom panels of the left column show the results for $\beta$, $b_\delta$, and $b_\eta$, respectively, while those of the right column do so for the perpendicular modes of \pthreed, the parallel modes of \pthreed, and \poned. Blue, orange, and red lines show the response of the previous quantities to a 5\% increase in $A_\mathrm{s}$, $\Omega_\mathrm{M}h^2$, and $\sigma_\mathrm{T}$, respectively, while green lines do so for a 1\% increase in $\bar{F}$.
}
\label{fig:sensitivity}
\end{figure*}

Small-scale \lya analyses use emulators to predict \poned as a function of cosmology and IGM physics \citep[e.g.,][]{cabayol-garcia2023NeuralNetworkEmulator}, while large-scale analyses use linear or perturbation theory models to predict \xithreed together with \lya linear bias parameters that need to be marginalized over. \forestflow provides a relationship between IGM physics and linear biases, enabling the use of \poned studies to inform three-dimensional analyses and vice versa.

We could use \forestflow to set constraints on $b_\delta$ and $b_\eta$ by fitting \poned measurements, and then use these constraints as priors in three-dimensional studies. As a result, we would break degeneracies between \lya linear bias parameters and cosmology, allowing us to measure the amplitude of linear density and velocity fluctuations, $\sigma_8(z)$ and $f \sigma_8(z)$, rather than $b_\delta \sigma_8$ and $b_\eta f \sigma_8$ like in traditional \lyaf analyses. To illustrate this application, we proceed to compare measurements of $b_\delta$ and $\beta\equiv b_\delta^{-1} b_\eta f$ from BAO analyses with \forestflow predictions for these parameters based on small-scale \poned analyses. The analysis of BAO in the \lyaf from the first data release of DESI yields $b_\delta=-0.108\pm0.005$ and $\beta=1.74\pm0.09$ at $z=2.33$ \citep{desicollaboration2024DESI2024IV}. On the other hand, \forestflow predicts $b_\delta=-0.118$ and $\beta=1.57$ at $z=2.33$ for a {\it Planck} cosmology when using as input the best-fitting constraints on IGM parameters from table 4 of \citet{emugp_Walther2019}, which were derived from high-resolution \poned measurements. The constraints on IGM parameters were derived using a \poned emulator trained on a suite of simulations with the same input cosmology and possibly slightly different definitions of IGM parameters relative to those used in this work. Nonetheless, \forestflow predictions and DESI measurements agree at the two sigma level, encouraging this new type of study.

In the left panels of Fig.~\ref{fig:sensitivity}, we display \forestflow predictions for the response of the \lya linear biases and $\beta$ to variations in cosmology and IGM physics. The response of $b_\delta$ to these changes is strong and has a different redshift dependence for cosmology and IGM parameters; therefore, we could use \forestflow to analyze \pthreed measurements from different redshifts to further break degeneracies between $b_\delta$ and $\sigma_8$. On the other hand, the response of $b_\eta$ to these changes is weak, and it is thus challenging to use this approach to break degeneracies between $b_\eta$ and $f \sigma_8$. Note that the response of the \lya linear biases and $\beta$ to $A_\mathrm{s}$ variations broadly agrees with measurements from simulations run while only varying $\sigma_8$ \citep{arinyo-i-prats2015NonlinearPowerSpectrum}.

Similarly, we could use measurements of linear bias parameters from three-dimensional analyses \citep{dumasdesbourboux2020CompletedSDSSIVExtended, desicollaboration2024DESI2024IV} to make predictions for IGM parameters, which could be used in \poned studies to break degeneracies between cosmology and IGM physics. In the right panels of Fig.~\ref{fig:sensitivity}, we display \forestflow predictions for the response of \pthreed and \poned to variations in cosmology and IGM physics. As we can see, the response of \poned to $A_\mathrm{s}$ and $\bar{F}$ variations is largely scale-independent down to $k_\parallel=1\iMpc$ where many other effects are at play, and thus these two parameters are largely degenerated. On the other hand, this is not the case for \pthreed; consequently, we could use information from \pthreed analyses to break degeneracies in \poned studies. Note that the response of \pthreed and \poned to $A_\mathrm{s}$, $\bar{F}$, and $\sigma_\mathrm{T}$ variations broadly agrees with measurements from simulations run varying only one of these parameters at a time \citep{mcdonald2003MeasurementCosmologicalGeometry, mcdonald2005LinearTheoryPower}.

We also observe that \pthreed and \poned respond significantly to variations in $\Omega_\mathrm{M}h^2$, suggesting that the \lya clustering is highly sensitive to the expansion and growth history. However, we find that variations in $A_\mathrm{s}$ and $n_\mathrm{s}$ can absorb the changes in \pthreed and \poned to the 2\% level, and completely do so for \pthreed at the pivot scale of the cosmological parameters of \forestflow, $k_\mathrm{p}=0.7\iMpc$. Furthermore, {\it Planck} measured $\Omega_\mathrm{M}h^2$ with 0.8\% precision \citet{planckcollaboration2020Planck2018Resultsa}, and $A_\mathrm{s}$ and $n_\mathrm{s}$ absorb 1\% variations in $\Omega_\mathrm{M}h^2$ to the $\simeq0.4\%$ level. This result supports the approach of not considering any cosmological parameter related to variations in the expansion of growth history as input for \forestflow (see also Sect.~\ref{sec:results_other}).


\subsection{Alcock-Paczy\'nski on mildly nonlinear scales}

Thanks to the increasing precision of galaxy surveys, there is a growing interest in extracting cosmological information from increasingly smaller scales in three-dimensional analyses. An avenue to do so is to analyze anisotropies in the correlation function \citet[AP test;][]{alcock1979EvolutionFreeTesta}, first proposed in the context of the \lyaf by \citet{1999ApJ...518...24M} and \citet{hui1999GeometricalTestCosmological}. Recently, \cite{cuceu2023ConstraintsCosmicExpansion} followed this approach to analyze \lyaf measurements from the Sloan Digital Sky Survey (SDSS) data release 16 \citep[DR16;][]{Ahumada2020_DR16}, yielding constraints on some cosmological parameters a factor of two tighter than those from BAO-only analyses. 

This study modeled three-dimensional correlations using linear theory, which restricted the range of scales analyzed to those larger than $25 \hMpc$. We could significantly extend the range of scales used in this type of analysis by modeling three-dimensional correlations using \forestflow. As a result, the constraining power of AP analyses would be much larger. Furthermore, we could use \forestflow to extract information from \poned analyses to reduce degeneracies between cosmology and the parameters describing \xithreed (see Sect.~\ref{sec:discussion_large_small}).


\subsection{Extending 3D analyses to the smallest scales}

\begin{figure}
    \centering
    \includegraphics[width=.9\columnwidth]{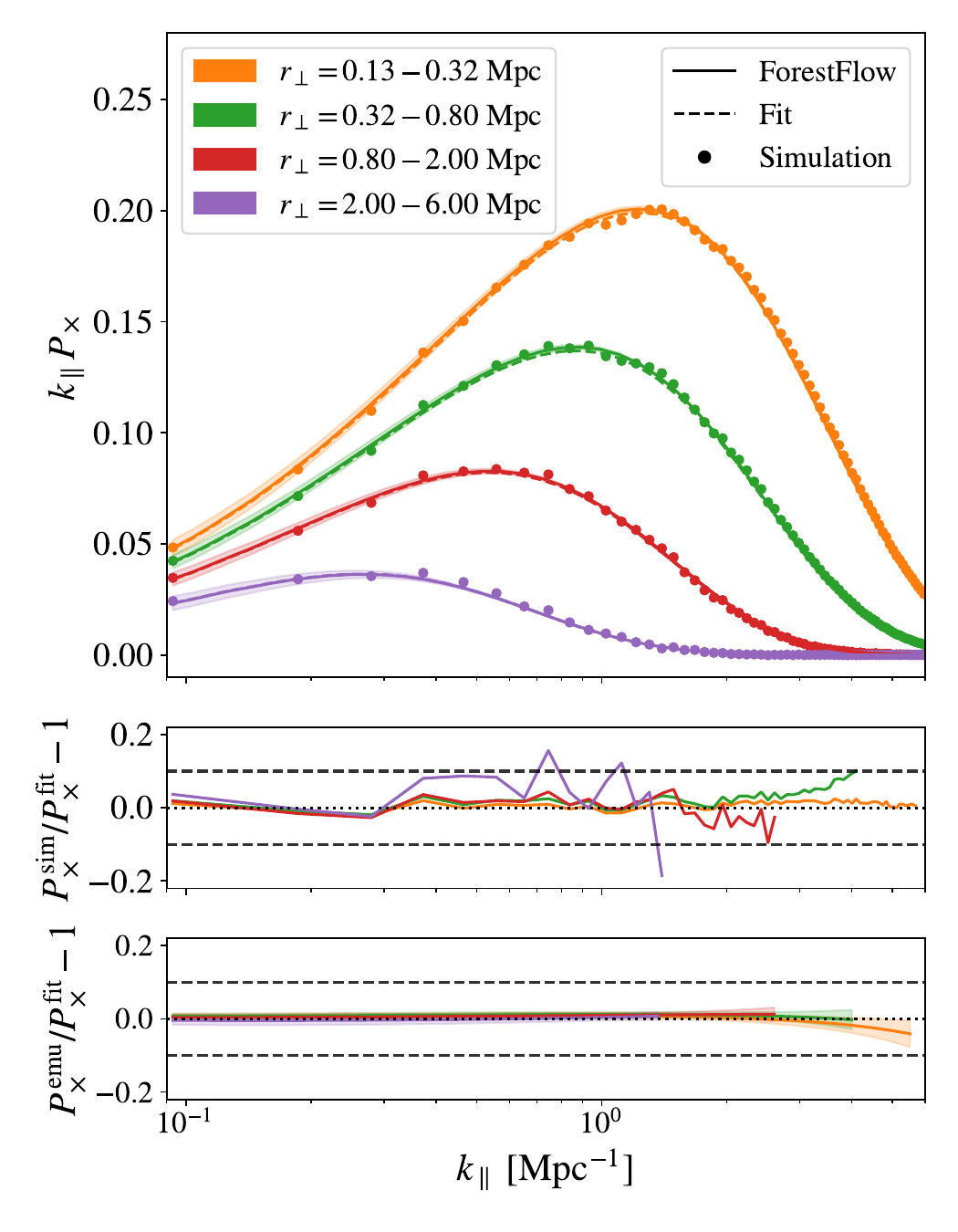}
    \caption{Accuracy of the parametric model and \forestflow in describing \pcross measurements from the \simcentral simulation at $z=3$. Dots show simulation measurements, dashed lines depict predictions from the best-fitting parametric model to \pthreed and \poned measurements, and solid lines and shaded areas display the average and 68\% credible interval of \forestflow predictions. The color of the lines indicates the results for different bins in transverse separation $r_\perp$. The middle panel shows the residual between simulation measurements and the best-fitting parametric model, while the bottom panel displays the residual between predictions from the parametric model and \forestflow. The performance of \forestflow in reproducing simulation measurements is similar to that of the best-fitting model.}
    \label{fig:Px_onesnap}
\end{figure}

The ultimate goal of \forestflow is to perform a joint analysis of one- and three-dimensional measurements from small to large scales. An interesting approach to do so is to measure the \lyaf cross-spectrum \citep[\pcross; e.g.,][]{hui1999GeometricalTestCosmological, fontribera2018HowEstimate3D}, which captures the correlation between one-dimensional Fourier modes from two neighboring quasars separated by a transverse separation ($r_\perp$). We can model this statistic by taking the inverse Fourier transform of \pthreed only along the perpendicular directions
\begin{align}
    P_{\times}(k_{\parallel}, r_\perp) &\equiv \frac{1}{(2\pi)^2}\int \mathrm{d}\boldsymbol{k}_\perp \, e^{i\,\boldsymbol{k}_\perp\cdot \boldsymbol{r}_\perp} \, P_\mathrm{3D}(k_\parallel, k_\perp) \nonumber\\ 
    &= \frac{1}{2\pi} \int_0^{\infty}  \mathrm{d}k_\perp\,  k_\perp\, J_0 (k_\perp r_\perp) \, P_\mathrm{3D}(k_\parallel, k_\perp) ~. \label{eq:pcross_bessel}
\end{align}
Comparing this equation with Eq.~\ref{eq:p1d}, it becomes clear that \poned is a special case of \pcross, corresponding to the limit where the transverse separation is zero.

In Sect.~\ref{sec:input_fitting}, we optimize the \pthreed model to describe measurements of \pthreed and \poned from the \lacehc simulations. Then, in Sect.~\ref{sec:forestflow}, we use the distribution of best-fitting parameters as the training set for \forestflow, which predicts the value of \pthreed model parameters as a function of cosmology and IGM physics. Even though neither the best-fitting model nor \forestflow use \pcross for their optimization, we can make predictions of \pcross for the two. To do so, we first estimate \pthreed using the value of the model parameters using Eq.~\ref{eq:p3d_model}, and then we integrate it using Eq.~\ref{eq:pcross_bessel}. We carry out the integration using the fast Hankel transform algorithm \texttt{FFTlog} \citep{Hamilton2000MNRAS.312..257H} implemented in the \texttt{hankl} package \citep{karamanis2021hankl}.

We use \pcross measurements from the simulations described in \S~\ref{sec:input_sims} to evaluate the accuracy of \forestflow for this statistic. We first define four bins in $r_\perp$, the transverse separation between skewers in configuration space, with edges 0.13, 0.32, 0.80, 2, and 6 Mpc. Then, we measure \pcross using all pairs of skewers with $r_\perp$ separation within the previous bins 
\begin{equation}
    P_\times(r_\perp, k_\parallel) = \bigg \langle \Re \Big[\tilde{\delta_\mathrm{i}}(k_\parallel) \tilde{\delta_\mathrm{j}}^*(k_\parallel)\Big]\bigg\rangle \,
\end{equation}
where $\tilde{\delta}_\mathrm{i}$ and $\tilde{\delta}^*_\mathrm{j}$ stand for the Fourier transform of a skewer $i$ and the complex conjugate of its partner $j$, respectively, the average $\langle\rangle$ includes all possible pairs in the bin without repetition or permutation, and $\Re$ indicates that we only use the real part of the expression between brackets because the average of the imaginary part is zero. The $r_\perp$ on the left-hand side denotes the effective center of the bin, accounting for the skewed distribution of $r_\perp$ within each bin: the number of skewers separated by a small distance $\mathrm{d}r_\perp$ is proportional to $r_\perp$, and therefore the effective center is not at the halfway point. To compute \pcross at the effective center, we perform the integration using ten sub-bins within each $r_\perp$ bin and calculate the average of these weighed by $r_\perp$.
    
In Fig.~\ref{fig:Px_onesnap}, we study the performance of \forestflow in reproducing \pcross measurements from the \simcentral simulation at $z=3$. Dots display simulation measurements, dashed lines the best-fitting model to \pthreed and \poned measurements from this simulation, and the solid lines \forestflow predictions. As we can see, \pcross decreases as the $r_\perp$ separation increases; this is because more distant sightlines are sampling increasingly uncorrelated regions. In the middle panel, we examine the accuracy of the best-fitting model in describing simulation measurements, finding that it is better than 10\% throughout all the scales shown. The performance of the model improves for smaller $r_\perp$ separations. This is likely because the fit's likelihood function (Eq.~\ref{eq:chi2}) considers \poned, which is equivalent to \pcross at $r_\perp=0$ separation, but not \pcross. The bottom panel illustrates the performance of \forestflow relative to the best-fitting model, providing an approximate assessment of its ability to reproduce the training data. \forestflow achieves an accuracy better than 5

Future studies could use \forestflow for extracting constraints on cosmology and IGM physics from the analysis of \pcross measurements \citep[e.g.,][]{Karim2023}. Nevertheless, as with \poned, these analyses would also require modeling multiple systematics affecting \lya measurements such as damped \lya systems, metal line contamination, and AGN feedback.


\section{Conclusions}
\label{sec:conclusions}

We present \forestflow, a novel framework for predicting \lya clustering from linear and nonlinear scales as a function of cosmology and IGM physics. \forestflow employs conditional normalizing flows to emulate the eight parameters of a physically motivated model for \lya clustering: the two linear \lya biases ($b_\delta$ and $b_\eta$) and six parameters that capture small-scale deviations of the three-dimensional flux power spectrum (\pthreed) from linear theory. By combining this model with a Boltzmann solver, \forestflow predicts \pthreed and any derived statistics, including the two-point correlation function (\xithreed, the primary statistic for large-scale analyses), the one-dimensional \lya flux power spectrum (\poned, central to small-scale studies), and the cross-spectrum (\pcross, a promising tool for full-scale analyses).

We trained the emulator using the best-fitting parameters of the physically motivated model to measurements from a suite of 30 fixed-and-paired cosmological hydrodynamical simulations spanning 11 redshifts equally spaced between $z=2$ and 4.5 \citep{Pedersen2021}. Despite the moderate size of these simulations, \forestflow achieves an accuracy of 3\% for \pthreed from linear scales to $k=5\iMpc$ and 1.5\% for \poned down to $k_\parallel=4\iMpc$. \forestflow uncertainties arise from three comparable sources: the size of the training simulations, their number, and the limited flexibility of the physically motivated model. However, we only evaluate the accuracy of \forestflow using simulations from our suite, and its performance may vary for higher-resolution simulations or those generated with other codes.

\forestflow demonstrates comparable performance for two extensions to the $\Lambda$CDM model --- massive neutrinos and curvature --- and ionization histories not included in the training set. This generalization is made possible by emulating the parameters of the physically motivated model as a function of the small-scale amplitude and slope of the linear power spectrum, the mean transmitted flux fraction, the amplitude and slope of the temperature-density relation, and the pressure smoothing scale \citep[see][]{pedersen2023CompressingCosmologicalInformation, cabayol-garcia2023NeuralNetworkEmulator}.

The release of \forestflow is particularly timely for \lyaf analyses with the ongoing Dark Energy Spectroscopic Instrument (DESI) survey. As discussed in Sect.~\ref{sec:discussion}, \forestflow enables a range of novel analyses with DESI data, including linking large- and small-scale studies and extending three-dimensional analyses to smaller scales. However, before applying \forestflow to cosmological inference, it will be necessary to model the effects of various astrophysical processes on \lya clustering, including metal contamination, damped \lya systems, and AGN feedback.


\begin{acknowledgements}
We thank Eric Armengaud, Roger de Belsunce, Andrei Cuceu, Vid Irsic, Ignasi P\'erez-R\`afols, and Michael Walther for their useful comments and helpful discussions. We also thank the referee for his detailed comments and suggestions. JCM, LCG, ML, and AFR acknowledge support from the European Union (ERC Consolidator Grant, COSMO-LYA, grant agreement 101044612). Views and opinions expressed are however those of the authors only and do not necessarily reflect those of the European Union or the European Research Council Executive Agency. Neither the European Union nor the granting authority can be held responsible for them. AFR acknowledges financial support from the Spanish Ministry of Science and Innovation under the Ramon y Cajal program (RYC-2018-025210) and the PGC2021-123012NB-C41 project. IFAE is partially funded by the CERCA program of the Generalitat de Catalunya. The analysis has been performed at Port d’Informaci\'o Cient\'ifica (PIC); we acknowledge the support provided by PIC in granting us access to their computing resources.

This material is based upon work supported by the U.S. Department of Energy (DOE), Office of Science, Office of High-Energy Physics, under Contract No. DE–AC02–05CH11231, and by the National Energy Research Scientific Computing Center, a DOE Office of Science User Facility under the same contract. Additional support for DESI was provided by the U.S. National Science Foundation (NSF), Division of Astronomical Sciences under Contract No. AST-0950945 to the NSF’s National Optical-Infrared Astronomy Research Laboratory; the Science and Technology Facilities Council of the United Kingdom; the Gordon and Betty Moore Foundation; the Heising-Simons Foundation; the French Alternative Energies and Atomic Energy Commission (CEA); the National Council of Humanities, Science and Technology of Mexico (CONAHCYT); the Ministry of Science and Innovation of Spain (MICINN), and by the DESI Member Institutions\footnote{\url{https://www.desi.lbl.gov/collaborating-institutions}}. Any opinions, findings, and conclusions or recommendations expressed in this material are those of the author(s) and do not necessarily reflect the views of the U. S. National Science Foundation, the U. S. Department of Energy, or any of the listed funding agencies. The authors are honored to be permitted to conduct scientific research on Iolkam Du’ag (Kitt Peak), a mountain with particular significance to the Tohono O’odham Nation.
\end{acknowledgements}

\section*{Data Availability}

\forestflow and all the notebooks used to generate the plots of this paper can be found in Github\footnote{\url{https://github.com/igmhub/ForestFlow}}, as well as all data points shown in the published graphs. The simulations utilized for training and testing the emulator are also publicly accessible\footnote{\url{https://github.com/igmhub/LaCE}}.



\bibliographystyle{aa_url}
\bibliography{forestflow} 

\begin{appendix}


\section{Cosmic variance}
\label{sec:cosmic_variance}

\begin{figure}
    \centering\includegraphics[width=\columnwidth]{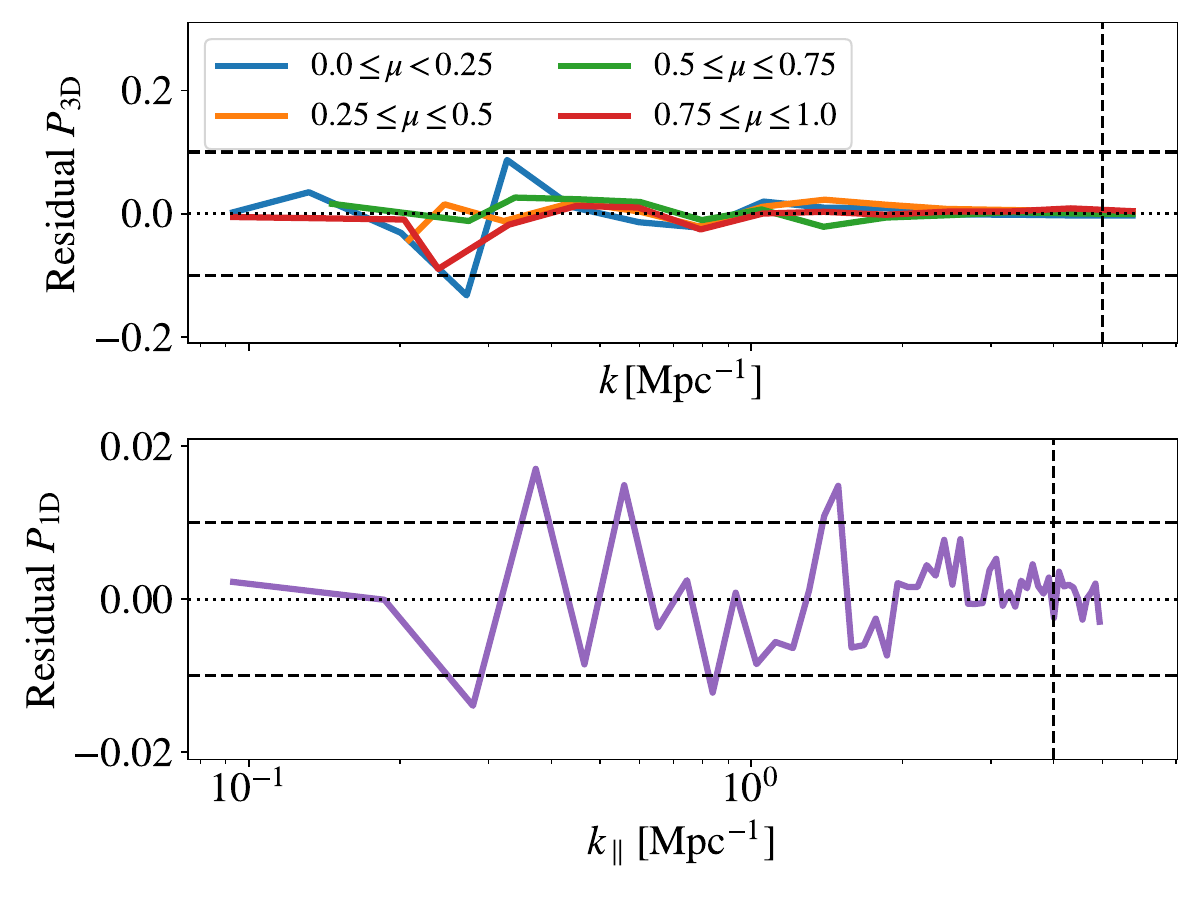}
    \caption{Impact of cosmic variance on \pthreed (top panel) and \poned (bottom panel) measurements from our simulations at $z=3$. Lines show the difference between measurements from the \simcentral and \simseed simulations, which only differ on their initial distribution of Fourier phases, divided by $\sqrt{2}$ times their average. Cosmic variance induces errors as large as 10\% on \pthreed for $k\simeq0.3\iMpc$, while these are on the order of $1\%$ for \poned.}
    \label{fig:cvar}
\end{figure}

\begin{figure}
\includegraphics[width=\columnwidth]{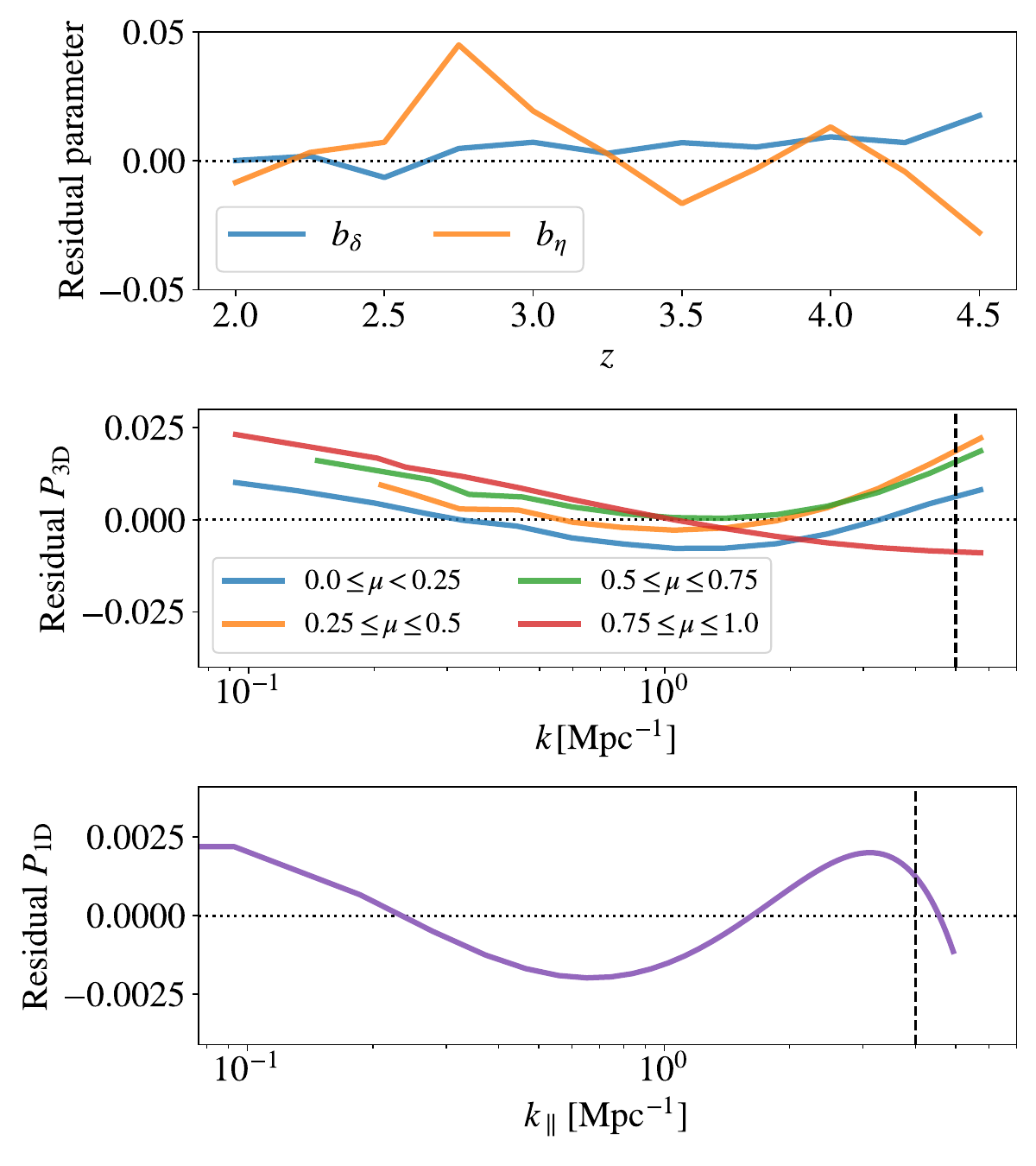}
\centering
\caption{Impact of cosmic variance on predictions from the parametric model. Lines show the difference between the best-fitting models to \pthreed and \poned measurements from the \simcentral and \simseed simulations, divided by $\sqrt{2}$ times the best-fitting model to their average. The top panel shows the results for the \lya linear biases ($b_\delta$ and $b_\eta$), while the middle and bottom panels display the results for \pthreed and \poned at $z=3$, respectively. The impact of cosmic variance on model predictions is approximately an order of magnitude smaller than on simulation measurements (see Fig.~\ref{fig:cvar}).}
\label{fig:cvar_fit}
\end{figure}

Throughout this work, we trained and tested \forestflow using simulations run employing the "fixed-and-paired" technique \citep{angulo2016CosmologicalNbodySimulations, pontzen2016InvertedInitialConditions}, which significantly reduces cosmic variance for the clustering of the \lyaf \citep{anderson2019CosmologicalHydrodynamicSimulations}. We could further mitigate the impact of cosmic variance by using control variates \citep{Kokron2022}, but this is outside the scope of the current work. The impact of cosmic variance on fixed-and-paired simulations is not straightforward \citep{maion2022fpvariance}, and thus we would ideally use multiple fixed-and-paired simulations with different initial distributions of Fourier phases to estimate the precision of measurements from our simulations. However, we only have two simulations with these properties: \simcentral and \simseed. In this section, we use these two simulations to estimate the impact of cosmic variance on simulation measurements and best-fitting models. It is crucial to acknowledge that our findings are subject to significant noise because we only have access to two independent realizations.

In Fig.~\ref{fig:cvar}, we show the difference between measurements from the \simcentral and \simseed simulations at $z=3$, normalized by $\sqrt{2}$ times their average\footnote{The factor $\sqrt{2}$ accounts for the noise estimate from a single simulation.}. The \simcentral and \simseed simulations differ only in their initial Fourier phase distributions, allowing their difference to isolate the effects of cosmic variance. Unlike traditional simulations, where cosmic variance for \pthreed scales inversely with the square root of the number of modes, this uncertainty peaks at $\simeq10\%$ around $k\simeq0.3\iMpc$ and decreases at both larger and smaller scales. This behavior can be explained as follows: at the largest scales, the fix-and-paired technique cancels cosmic variance for linear density modes, reducing variance. At intermediate scales, nonlinear evolution, particularly mode coupling, reintroduces cosmic variance, increasing the uncertainty. At smaller scales, the increasing number of modes leads to a decrease in cosmic variance, similar to trends observed in traditional simulations.

Thus, cosmic variance limits our ability to accurately evaluate both the parametric model and \forestflow using our simulations. To mitigate its impact, we restricted the analysis of \pthreed performance to scales $k > 0.5\iMpc$ in the main results. In contrast, cosmic variance has a much smaller effect on \poned, contributing only about $1.5\%$ uncertainty at $k_\parallel < 2\iMpc$. This smaller effect allowed us to include all scales in \poned tests without concern. To more precisely quantify the impact of cosmic variance on simulation measurements, we computed the standard deviation of the results shown in Fig.~\ref{fig:cvar} across redshift. We did it within the intervals $0.5 < k[\iMpc] < 5$ for \pthreed and $0.09 < k_\parallel[\iMpc] < 4$ for \poned, based on the scales discussed earlier and those used for fitting the \pthreed model in Sect.~\ref{sec:input_fitting}. We found that the average impact of cosmic variance is $1.3\%$ for \pthreed and $0.5\%$ for \poned.

We expect the impact of cosmic variance on the best-fitting model to \pthreed and \poned measurements to be weaker than on simulation measurements, as multiple \pthreed and \poned bins collectively constrain the eight parameters of the model. In Fig.~\ref{fig:cvar_fit}, we show the difference between the best-fitting models to the \simcentral and \simseed simulations, normalized by $\sqrt{2}$ times the best-fitting model to their average. The top panel displays results for the two \lya\ linear biases, $b_\delta$ and $b_\eta$. The standard deviation of the differences is $0.6\%$ and $1.8\%$ for $b_\delta$ and $b_\eta$, respectively, demonstrating that we can measure the two \lya\ linear biases with percent-level accuracy using our simulations. This precision is achievable because the combination of small and large scales in the fits helps break degeneracies between the two linear biases and the other six model parameters. 

By propagating these uncertainties to the behavior of \pthreed on linear scales, we find that the impact of cosmic variance on perpendicular and parallel modes is $1.2\%$ and $1.8\%$, respectively. In the middle and bottom panels of Fig.~\ref{fig:cvar_fit}, we analyze the influence of cosmic variance on model predictions for \pthreed and \poned. The overall impact of cosmic variance on \pthreed and \poned predictions is $0.8\%$ and $0.1\%$, respectively, confirming that the best-fitting model is less sensitive to cosmic variance than individual simulation measurements. Consequently, \forestflow is more robust against cosmic variance than emulators that predict the power spectrum at fixed $k$-bins.


\section{Validation of uncertainties predicted for \pthreed and \poned}
\label{sec:uncertainty_validation}

\begin{figure}
    \centering\includegraphics[width=\columnwidth]{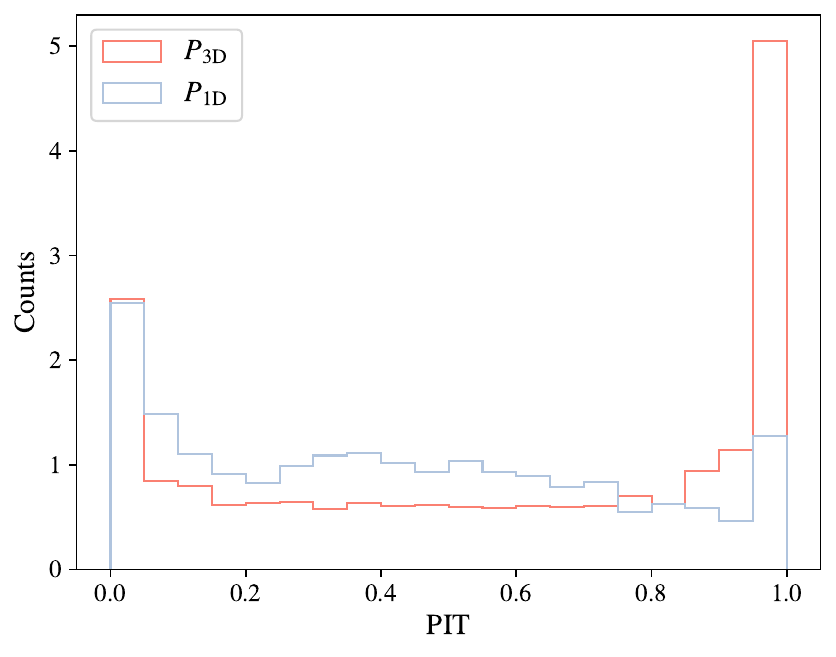}
    \caption{PIT distribution for \poned (blue) and \pthreed (red). This plot validates the uncertainties predicted by \forestflow across \lacehc simulations via a leave-simulation-out approach. The PIT distribution is approximately uniform, indicating well-calibrated uncertainties for most samples, while the peaks at the edges indicate underestimated uncertainties for some samples.}
    \label{fig:PIT}
\end{figure}

Normalizing flows predict the full posterior distribution of the target data rather than only their mean like fully connected neural networks or their mean and width like Mixture Density Networks \citep[see][for some applications in cosmology]{ramachandra2022MachineLearningSynthetic, cabayol-garcia2023NeuralNetworkEmulator}. This is achieved through multiple sampling iterations from the target latent distribution, an eight-dimensional Gaussian in our case. In \forestflow, each sampled realization of the \pthreed model parameters is propagated to generate predictions for \pthreed and \poned (see Sect.~\ref{sec:forestflow_NF}), producing a covariance matrix for these statistics. In this appendix, we validate its diagonal elements. Note that well-calibrated uncertainties are critical for future uses of \forestflow such as cosmology inference.

We validated the uncertainty in \pthreed and \poned predictions using the Probability Integral Transform test (PIT), which is the value of the cumulative distribution function (CDF) of a distribution evaluated at the ground-truth value $z_{\rm t}$
\begin{equation}
    \rm{PIT} = \mathrm{CDF}[p,\,z_{\rm t}]=\int_{-\infty}^{z_{\rm t}} p(z) \mathrm{d}z\, ,
    \label{eq:pit} 
\end{equation} 
where $p$ is in our case the distribution of \forestflow predictions for \pthreed or \poned and $z_{\rm t}$ stands for measurements of these statistics from the simulations. A model that displays a well-calibrated uncertainty distribution yields PIT values that are uniformly distributed between zero and one. This indicates that the observed outcomes have an equal likelihood of falling at any point along the predicted CDF. In contrast, an excess of values close to zero or one indicates that the width of the distribution is underestimated.

In Fig.~\ref{fig:PIT}, we display a PIT test produced using all the \lacehc simulations via a leave-simulation-out approach (see Sect.~\ref{sec:results_statistics}). This process validates average predictions and uncertainties against simulations excluded in the training process. The red and blue lines display the results for \pthreed and \poned, respectively, which were generated by combining results from different scales and redshifts. As we can see, the PIT distribution is approximately uniform for the two statistics but it presents peaks at the low and high ends, indicating underestimated uncertainties for some samples. The cause behind this feature is unclear and it demands further investigation beyond the scope of this project.

\end{appendix}
\end{document}